\documentclass[pdflatex,sn-mathphys-num]{sn-jnl}

\usepackage{physics}
\usepackage{graphicx}%
\usepackage{multirow}%
\usepackage{amsmath,amssymb,amsfonts}%
\usepackage{amsthm}%
\usepackage{mathrsfs}%
\usepackage[title]{appendix}%
\usepackage{xcolor}%
\usepackage{textcomp}%
\usepackage{manyfoot}%
\usepackage{booktabs}%
\usepackage{algorithm}%
\usepackage{algorithmicx}%
\usepackage{algpseudocode}%
\usepackage{listings}%
\usepackage{threeparttable}
\usepackage{makecell}
\usepackage{array}
\usepackage{siunitx}
\usepackage[utf8]{inputenc}
\usepackage[T1]{fontenc}

\theoremstyle{thmstyleone}%
\theoremstyle{thmstyletwo}%

\theoremstyle{thmstylethree}%

\begin{document}

\title[Article Title]{Learning spatially structured open quantum dynamics with regional-attention transformers}


\author*[1]{\fnm{Dounan} \sur{Du}}\email{dounan.du@stonybrook.edu}


\author*[1,2]{\fnm{Eden} \sur{Figueroa}}\email{eden.figueroa@stonybrook.edu}

\affil*[1]{\orgdiv{Department of Physics and Astronomy}, \orgname{Stony Brook University}, \orgaddress{\city{Stony Brook}, \postcode{11794-3800}, \state{NY}, \country{USA}}}

\affil*[2]{\orgname{Brookhaven National Laboratory}, \orgaddress{\city{Upton}, \postcode{11973-5000}, \state{NY}, \country{USA}}}


\abstract{Simulating the dynamics of open quantum systems with spatial structure and external control is an important challenge in quantum information science. Classical numerical solvers for such systems require integrating coupled master and field equations, which is computationally demanding for simulation and optimization tasks and often precluding real-time use in network-scale simulations or feedback control. We introduce a regional attention-based neural architecture that learns the spatiotemporal dynamics of structured open quantum systems. The model incorporates translational invariance of physical laws as an inductive bias to achieve scalable complexity, and supports conditioning on time-dependent global control parameters. We demonstrate learning on two representative systems: a driven dissipative single qubit and an electromagnetically induced transparency (EIT) quantum memory. The model achieves high predictive fidelity under both in-distribution and out-of-distribution control protocols, and provides substantial acceleration up to three orders of magnitude over numerical solvers. These results demonstrate that the architecture establishes a general surrogate modeling framework for spatially structured open quantum dynamics, with immediate relevance to large-scale quantum network simulation, quantum repeater and protocol design, real-time experimental optimization, and scalable device modeling across diverse light–matter platforms.}

\maketitle

\section{Introduction}\label{sec1}

Open quantum systems are fundamental to the operation of quantum memories, network nodes, repeaters, and light–matter interfaces across quantum information science. These devices are realized across diverse platforms including cold atom ensembles\cite{radnaev2010quantum, wang2019efficient, vernaz2018highly}, atom arrays\cite{chang2018colloquium, reitz2022cooperative, masson2022universality}, room-temperature vapor-based devices\cite{willis2009four, ogden2019quasisimultons}, and engineered light–matter interfaces in waveguide\cite{chang2018colloquium, corzo2019waveguide} or cavity QED\cite{kumar2023quantum}. In many of these platforms, spatial propagation and time-dependent driving fields fundamentally shape the dynamics, giving rise to rich interplay between coherent quantum evolution, dissipative processes, and structured spatiotemporal behavior. A paradigmatic example is the electromagnetically induced transparency (EIT)-based quantum memory\cite{fleischhauer2000dark, fleischhauer2002quantum, fleischhauer2005electromagnetically, lvovsky2009optical, namazi2017ultralow, chaneliere2005storage, gera2024hong}, where probe pulses propagate through a spatially extended atomic medium under a time-dependent control field. Such systems are usually modeled by quantum master equation coupled with field propagation equations\cite{fleischhauer2002quantum}. Accurate simulation over extended spatiotemporal domains and under time dependent control protocols is computationally intensive\cite{ogden2019quasisimultons, potvliege2025coombe}, particularly when required repeatedly for optimization, network-scale modeling, or real-time experimental feedback.

Deep learning is increasingly being explored as a tool for assisting study of physical dynamic systems\cite{raissi2019physics, bar2019learning, kovachki2023neural}, as well as for assisting quantum information experiments\cite{lin2025ai}. Prior work has applied neural networks to simulate Lindblad evolution, quantum trajectories, and operator dynamics in relatively low-dimensional or few-body settings\cite{von2022self, choi2022learning, viteritti2025transformer, zhang2025neural}. Among deep learning approaches, transformer architectures, originally developed for language and vision tasks \cite{vaswani2017attention, dosovitskiy2020image}, have shown strong performance in learning physics systems dynamics governed by partial differential equations \cite{geneva2022transformers, mccabe2024multiple}, and have begun to be applied to quantum models\cite{sprague2024variational}. However, most existing studies focus on temporally localized or low-dimensional quantum systems, and do not address quantum systems with spatial propagation, global control protocols, and decoherence. Moreover, the quadratic scaling of standard self-attention mechanisms in sequence length poses a challenge for modeling spatial-temporal quantum systems with fine resolution. Recent advances in scalable attention mechanisms from the computer vision and geoscience communities, including axial attention \cite{ho2019axial}, Swin Transformers \cite{liu2021swin, liu2022swin}, and Earthformer \cite{gao2022earthformer} offer potential architectural solutions, but their utility in modeling control-driven, dissipative quantum systems remains largely unexplored.

In this work, we propose a physics-informed regional transformer architecture designed to efficiently learn the dynamics of structured open quantum systems under external driving. The architecture is based on regional attention, which exploits translation invariance of the physical laws as an inductive bias to achieve scalable complexity. The architecture encodes local density matrix with build-in Hermiticity, and employs a causal decoder-only structure for autoregressive, physically consistent generation. It further supports conditioning on time-dependent global parameters, allowing it to capture how external control fields drive the system's evolution. 

We evaluate the architecture on (i) a driven dissipative qubit and (ii) a spatially extended EIT quantum memory, benchmarking fidelity, physical constraint preservation, and experimentally relevant observables (readout delay, pulse energy) under both in-distribution and out-of-distribution control parameters, with and without decoherence. In both cases, the model achieves high fidelity and robust generalization while providing up to ~1485x acceleration over numerical solvers on modern GPUs. Together, these results highlight a general surrogate modeling framework for structured open quantum dynamics, with potential applications in large-scale quantum network simulation, real-time experimental feedback, and scalable quantum information device optimization.

\section{Results}

\subsection{Problem Setup}
\begin{figure*}
\includegraphics[width=1\textwidth]{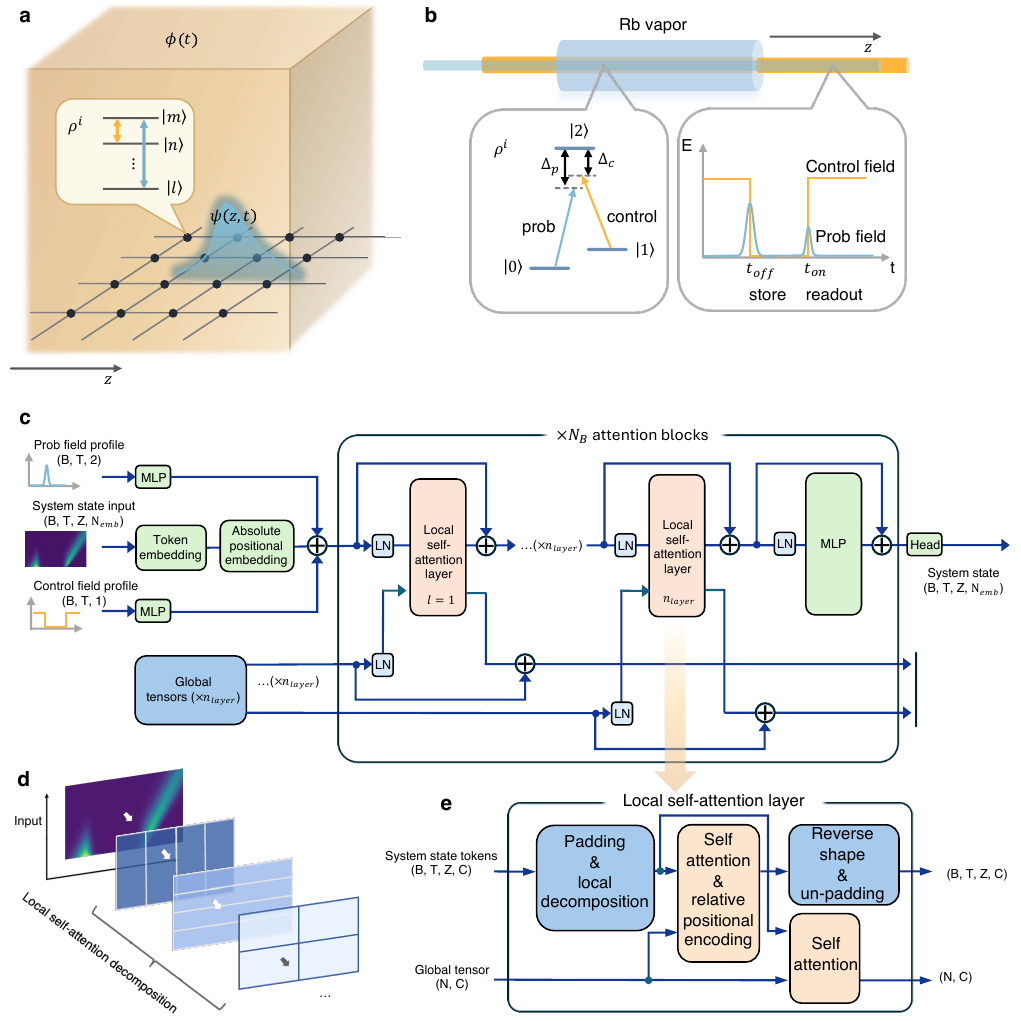}
\caption{\label{fig:main} \textbf{The problem setup and architectural details.} \textbf{a}, The problem setup. An spatial structured open quantum system is subject to a global field $\phi(t)$ and a propagating field $\psi(r_i, t)$. \textbf{b}, The EIT quantum memory setup. The control field and prob field are co-propagating in the $z$ direction in a Rb vapor cell. \textbf{c}, The Quformer architecture. \textbf{d}, The communication channel I. Different decomposition configurations are layered in an cyclic pattern in an attention block. \textbf{e}, The local self-attention layer scheme.}
\end{figure*}

We confine our interest in open quantum systems with a spatial structure(Fig.~\ref{fig:main}a), for example, a grid or a lattice. Each site may also coupled to a global time dependent control field $\phi(t)$, and a propagating field $\psi(r_i, t)$ satisfied by some propagation equation. The system Hamiltonian has the general form

\begin{equation}
    \begin{aligned}
        \hat{H} &= \sum_{i}\sum_{l}\epsilon_l\hat{\sigma}_l^i+\phi(t)\sum_i\sum_{lm}\hat{\sigma}_{lm}^i+\sum_i\sum_{lm}\psi(r_i, t)\hat{\sigma}_{lm}^i
    \end{aligned}
\end{equation}
where $\hat{\sigma}_l^i = \ket{l}\bra{l}$ and $\hat{\sigma}_{lm}^i = \ket{l}\bra{m}$ at site $i$. We further confine the system environment interaction under the Born–Markov approximation. The system evolution is thus governed by the quantum master equation
\begin{equation}
    \begin{aligned}
        \frac{d\rho^i}{dt} &= -i[H, \rho^i] + \sum_j \gamma_j \left( L_j \rho^i L_j^\dagger - \frac{1}{2} \left\{ L_j^\dagger L_j, \rho^i \right\} \right).
    \end{aligned}
\end{equation}
The model is often seen in quantum information applications involving light matter interaction.

\subsection{Model Design and Architecture}

We model the spatiotemporal evolution of structured open quantum systems on a discretized domain. The system is defined on a uniform spacetime grid, where each point $(r_i,t)$ encodes the local quantum state $\rho^i(r_i, t)$ and the propagating field $\psi(r_i, t)$. These quantities are combined into a state token, which serves as the fundamental unit for learning and prediction. The full system trajectory is represented as a sequence of spatial frames evolving over time. Internally, this corresponds to an array of shape (T, X, Y, Z, C), where T, X, Y, Z index time and spatial dimensions, and C denotes the token embedding dimension. The learning task is to predict future frames of the system evolution given only a limited number of initial observations.

Applying standard transformer architectures directly to structured quantum dynamics leads to a severe computational bottleneck. In these architectures, self-attention is computed across all token pairs, resulting in quadratic scaling with respect to the total number of tokens. For a four-dimensional spacetime domain, the attention complexity grows as $O(N_x^2N_y^2N_z^2N_t^2)$, where $N_i$ denotes the number of grid points along each dimension. This scaling rapidly becomes intractable for high-resolution grids, even for modest physical domains. To overcome this, we introduce a model architecture that exploits the translational invariance of the system equation of motion as an inductive bias. The evolution at different spacetime points is governed by the same local dynamical laws, expressed by Eqs. (2) and the coupled propagation equation of the field $\psi(r_i, t)$. This symmetry motivates a regional decomposition of the spacetime domain into fixed-size, non-overlapping subregions, where self-attention is applied locally. The attention weights learned within one subregion can then be shared across all others, enabling efficient and scalable modeling of global dynamics. To maintain casual connection across subregions, we incorporate communication channels that exchange boundary information, allowing the model to recover long-range interactions while preserving computational efficiency.

We embed a weak physics-informed constraint into the token representation by explicitly enforcing Hermiticity of the density matrix at each grid point. The complex-valued density matrix $\rho_{ij}(r_i, t)$ at every grid point is mapped to a real-valued matrix representation $\rho'_{ij}(r_i, t)$ by preserving diagonal elements and separating the real and imaginary components of off-diagonal terms:
\begin{equation}
    \begin{aligned}
        \rho'_{ij} = 
            \begin{cases}
            \rho_{ii}, & i = j, \\[8pt]
            \dfrac{1}{2} \left( \rho_{ij} + \rho_{ji} \right), & i < j, \\[8pt]
            \dfrac{1}{2\mathrm{i}} \left( \rho_{ji} - \rho_{ij} \right), & i > j.
            \end{cases}
    \end{aligned}
\end{equation}
The transformation matrix is then vectorized to form the quantum-state component of the token representation. To incorporate the propagating field, we concatenate the real and imaginary parts of the local field $\psi(r_i, t)$ with the vectorized density matrix $\mathbf{v}_i = \textbf{Cat}\{\mathrm{vec}(\rho'_{ij}), \mathrm{Re}[\psi(r_i, t)], \mathrm{Im}[\psi(r_i, t)]\}$, forming the complete input token at grid point $(r_i, t)$. This representation naturally aligns with the real-valued input requirements of deep learning architectures while preserving all information encoded in the original quantum state and field.



The regional decomposition is done by decomposing the full region into non-overlapping local region of shape (t, x, y, z, C): $(B, T, X, Y, Z, C) \rightarrow (B,\  \frac{T}{t}\cdot\frac{X}{x}\cdot\frac{Y}{y}\cdot\frac{Z}{z},\  t,\  x,\  y,\ z,\ C)$, where B is the batch size.  Different local regions are then treated as a new batch dimension of size $\frac{T}{t}\cdot\frac{x}{x}\cdot\frac{Y}{y}\cdot\frac{Z}{z}$. A self-attention layer is then performed over the flattened local region. The decomposition brings the attention complexity from $O(T^2X^2Y^2Z^2)$ to $O(TXYZ\times txyz)$ with a linear scale of the local region size. The regional decomposition ensures within different regions the system state evolves under the same token relations by sharing the same $W_Q$, $W_K$, $W_V$ matrix. However, exchange of information among different local regions are required to fully describe the system evolution over the full region of interest. In the architecture we employed two distinct communication channels among local regions.

\textbf{Communication channel I}: Alternating local region definition. In this channel, we employ $n_{layer}$ different self-attention layers(Fig.~\ref{fig:main}e) in an cyclic pattern within a self-attention block. Each layer applies a different local region decomposition configuration(Fig.~\ref{fig:main}d). The key idea is boundaries of local regions in one type of decomposition should be included(or partially included) within another type of decomposition in the next attention layer, thus two neighbor local regions in layer $l$ can exchange information in layer $l+1$. Formally, let $\Omega \subset \mathbb{R}^d$ be the overall region of interest. For each self-attention layer $l$ (with $l = 1, \dots, n_{\text{layer}}$), we partition $\Omega$ into non-overlapping local regions that are all congruent to a fixed template region $S \subset \mathbb{R}^d$. That is, for each $l$, there exists an index set $I_l$ and translation vectors $\{t^{(l)}_i \in \mathbb{R}^d : i \in I_l\}$ such that $R^{(l)}_i = S + t^{(l)}_i, \quad \forall\, i \in I_l$, and $\Omega = \bigcup_{i \in I_l} R^{(l)}_i \quad \text{(with the union being disjoint)}$. Denote by $\partial R^{(l)}_i$ the boundary tokens of the region $R^{(l)}_i$. The design requirement is for every $l\in\{1,\dots,n_{\text{layer}}\}$ and every $i\in I_l$, there exists a $j\in I_{l+1}$ such that $\partial R^{(l)}_i \cap R^{(l+1)}_j \neq \varnothing$.

\textbf{Communication channel II}: Data bus from global tensors. Similar to the global vectors from weather forecasting model Earthformer\cite{gao2022earthformer}, we extend the local region self-attention to include a global tensor $G$. The tokens within each local region attend not only themselves but also the global tensor
\begin{equation}
\begin{aligned}
    X^i_{local} = \mathrm{softmax}\left( \frac{(X^i_{local} W_Q)(\textbf{Cat}(G,X^i_{local}) W_K)^\top}{\sqrt{d_k}} \right) (\textbf{Cat}(G,X^i_{local}) W_V),
\end{aligned}
\end{equation}
and the global tensor updates by attending the full region of interest to aggregate cross-region information
\begin{equation}
\begin{aligned}
    G = \mathrm{softmax}\left( \frac{(G W'_Q)(X_{full} W'_K)^\top}{\sqrt{d_k}} \right) (X_{full} W'_V).
\end{aligned}
\end{equation}
For each decomposition corresponding to a layer $l$ within a self-attention block, we assign a global tensor $G^l$ associate with it. The same global tensor is shared across different self-attention blocks with corresponding layers. Global tensors are served as a data bus, connecting distributed local regions $R^{(l)}_i$.


In the probelm setup the system evolution is subject to two fields with distinct roles: a global control field and a propagating field. The control field $\phi(t)$, which is uniform in space but varies in time, is encoded using a multi-layer perceptron (MLP) and is embedded into each system token, analogous to learned positional encodings, enabling the model to condition the local dynamics on the global control profile. In contrast, the propagating field  $\psi(r_i, t)$ serves as a time dependent boundary input. We encode its time-dependent profile using a separate MLP and inject the result into the boundary tokens along the entire temporal axis. This dual-field embedding scheme allows the model to incorporate both global driving effects and time dependent boundary conditions, mirroring their roles in the underlying physical equations.

To reflect the causal structure of system evolution, we adopt a decoder-only architecture that generates the system dynamics in an autoregressive, frame-by-frame manner(Fig.~\ref{fig:main}c). The model is composed of $N_B$ stacked attention blocks, each structured by a cyclic regional decomposition scheme. Absolute positional encodings are applied to all tokens before decomposition to retain global temporal and spatial context. Within each subregion, we apply relative positional encodings allowing the model to learn local attention patterns while maintaining local spatial information. This architecture enables the model to iteratively generate future states conditioned on prior states, aligning the inference process with the physical evolution of the system.

\subsection{Controlled Single Qubit Dissipative Rabi Oscillation}

\begin{figure*}
\includegraphics[width=\linewidth]{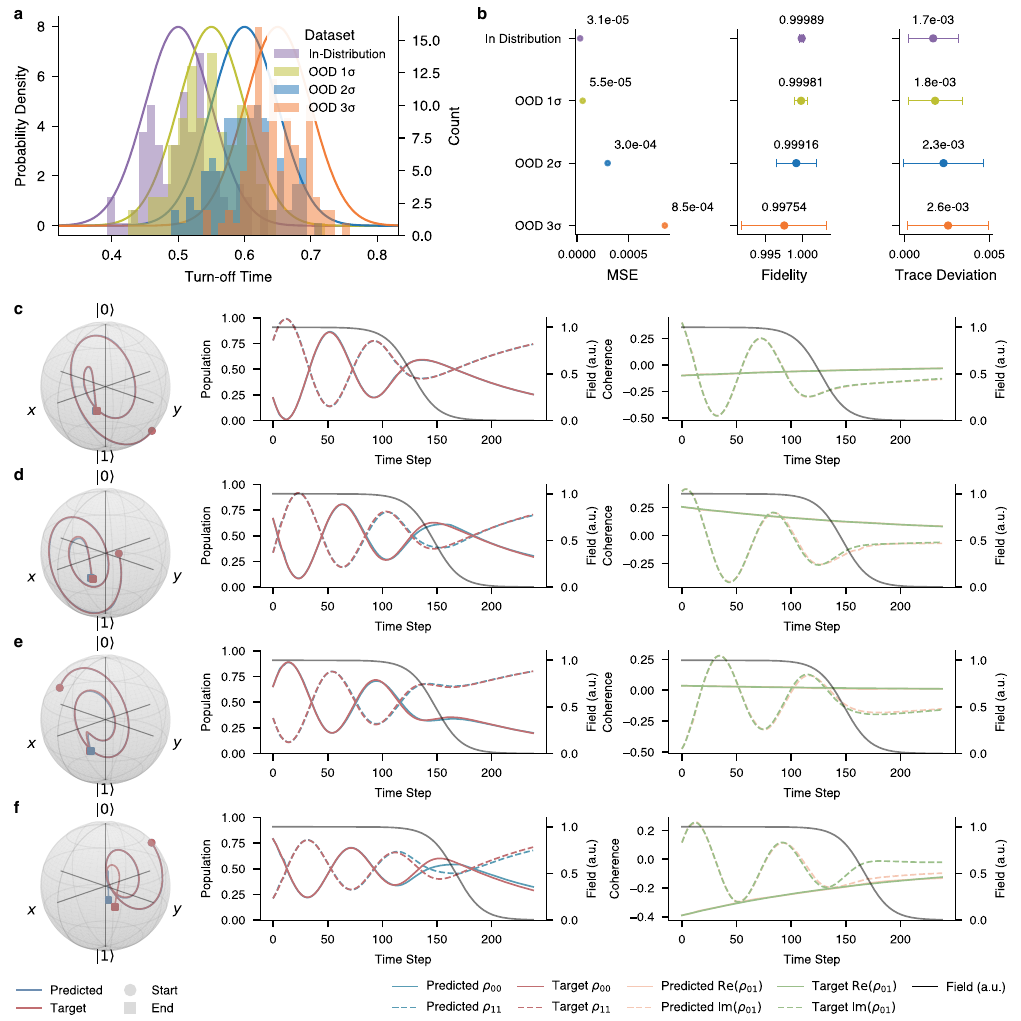}
\caption{\label{fig:qubit learning}\textbf{Single qubit learning evaluation.} \textbf{a}, Distribution of evaluation datasets over the driving field turn-off time $t_{off}$. Out-of-distribution (OOD) test sets are constructed by shifting the mean of the training distribution by $1\sigma$ (OOD1), $2\sigma$ (OOD2), and $3\sigma$ (OOD3). \textbf{b}, Model evaluation metrics (mean-squared error, fidelity, and trace deviation) across in-distribution (ID) and OOD datasets. Data points represent mean values, and error bars indicate one standard deviation. \textbf{c-f}, Representative model predictions compared with target trajectories visualized on the Bloch sphere and as density matrix elements trajectories. Results are shown for \textbf{c}, in-distribution, \textbf{d}, OOD1, \textbf{e}, OOD2, and \textbf{f}, OOD3 datasets, respectively. Deviations between predictions and targets become more pronounced at higher $\sigma$-shifts, particularly after the driving field is turned off.}
\end{figure*}

We first consider a two-level system undergoing dissipative Rabi oscillations driven by a time-dependent control field. This minimal open quantum system is confined to a single spatial point, and the system domain consists solely of a one-dimensional time axis. In this setting, the local spatial decomposition becomes trivial, and the regional attention mechanism naturally reduces to standard self-attention over the temporal sequence. As such, the single-qubit model provides a structurally simplified case that enables us to validate both the quantum state embedding scheme and the representation of global time-dependent control fields.

In the data-generating process, we fix the amplitude of the driving field and vary both the initial quantum state and the control-field turn-off time. The initial state is sampled from a Gaussian distribution over the superposition coefficients of $\ket{0}$ and $\ket{0}$, while the turn-off time $t_{off}$ is drawn from a separate Gaussian prior(Fig.~\ref{fig:qubit learning}a). To evaluate the model’s extrapolation performance, we construct out-of-distribution (OOD) test sets by shifting the mean of the $t_{off}$ distribution by $1\sigma$, $2\sigma$, and $3\sigma$, while keeping the initial state data-generating distribution fixed. This one-dimensional extrapolation protocol isolates the model’s generalization behavior along a single experimentally relevant axis and will later be applied in the context of a spatially extended quantum memory.

We evaluate the model’s performance using three complementary metrics: the mean-squared error (MSE), state fidelity, and trace deviation. The MSE measures the average prediction error across the full density matrix trajectories. Fidelity quantifies state similarity between predicted and target states, providing a direct tomography-based quantum state evaluation. The trace deviation assesses how accurately the model preserves the density matrix’s unit-trace constraint, a physical requirement not explicitly enforced in model architecture. On in-distribution test data, the model achieves high accuracy with an average MSE of $3.1\times10^{-5}$ and fidelity of $0.999894 \pm 0.000277$. Under distributional shift, the model exhibits smooth degradation: the MSE increases to $3.0\times10^{-4}$ at $2\sigma$ and $8.5\times10^{-4}$ at $3\sigma$, while fidelity remain above 0.997 across all test OOD sets (Fig.~\ref{fig:qubit learning}b). Trace deviation remains consistently low and stable, indicating effective structural learning of physical constraints.

Qualitative analysis of Bloch-sphere trajectories and density matrix elements trajectories further supports these findings, demonstrating that the model maintains the correct geometric structure of quantum state evolution up to $2\sigma$ shifts, with systematic deviations becoming visually apparent at $2\sigma$ shifts, particularly in the post-control field interval (Fig.~\ref{fig:qubit learning}c–f). These observations indicate that the model captures essential dynamical features, though its predictive precision is progressively limited under stronger extrapolation. Taken together, the results demonstrate controlled and physically consistent generalization within a clearly defined range of parameter shifts.

The single-qubit results demonstrate that the model accurately captures the dissipative Rabi dynamics and provides stable predictions under controlled extrapolation of the driving field’s turn-off time, within the tested range of $3\sigma$. These findings establish an initial validation of our token embedding and field encoding architecture in a minimal quantum setting and serve as a baseline for investigating its performance in more complex, spatially structured quantum systems, as explored next in the quantum memory application.

\subsection{Learning EIT quantum memory dynamics}


We consider a non-trivial example of a EIT-based quantum memory in Rubidium vapor cell(Fig.~\ref{fig:main}b). The atoms can be modeled as a three-level $\Lambda$-type system. The system Hamiltonian in rotating frame is 
\begin{equation}
\begin{aligned}
        \hat{\tilde{H}} &= \sum_i\hbar[ -\Delta_p \hat{\sigma}^i_3 + (-\Delta_p+\Delta_c) \hat{\sigma}^i_2]\\
    &+\sum_i \hbar[\frac{\Omega_c}{2} \hat{\sigma}_{23} + \frac{\Omega_p(z_i)}{2} \hat{\sigma}_{13} + h.c.].
\end{aligned}
\end{equation}
 We assume that the strong control field strength, hence the $\Omega_c$ is a global z-independent parameter. The sum is over all atoms inside the control and prob beam mode. We also assume the control and prob beam are co-propagating along the $z$ axis, satisfying the phase conservation. The rotating frame atom density matrix $\tilde{\rho}(z,t)$ is governed by the Master equation 
\begin{equation}
\begin{aligned}
\frac{d\tilde{\rho}(z,t)}{dt} &= -\frac{i}{\hbar} \comm{\tilde{H}}{\tilde{\rho}(z,t)} + \sum_k \left( L_k \tilde{\rho}(z,t) L_k^\dagger - \frac{1}{2} \acomm{L_k^\dagger L_k}{\tilde{\rho}(z,t)} \right).
\end{aligned}
\end{equation}
In the semi-classical model, the prob field is treated as a small classical field $E_p(z, t) = \frac{1}{2}\mathcal{E}_p(z, t)\ exp[-i(\nu t -kz+\phi_{z, t}) + c.c.]$ with Rabi frequency $\Omega_p=-\bra{1}\hat{d}\ket3\cdot \hat{\epsilon}\mathcal{E}_p/\hbar$. The prob field propagate under the propagation equation
\begin{equation}
\begin{aligned}
    (\frac{1}{c}\frac{\partial}{\partial t}+\frac{\partial}{\partial z})\mathcal{E}_p(z, t) &= \frac{ik}{\epsilon_0}Nd_{13}\tilde{\rho}_{31}(z, t)
\end{aligned}
\end{equation}
where $N$ is the atomic density inside the prob beam mode, $d_{13}$ is the dipole matrix element corresponding to transition $\ket{1}\rightarrow\ket{3}$. Equation (7) and (8) together constitutes a coupled equation that governs the system evolution over the optical mode region.

\begin{figure*}
\includegraphics[width=\linewidth]{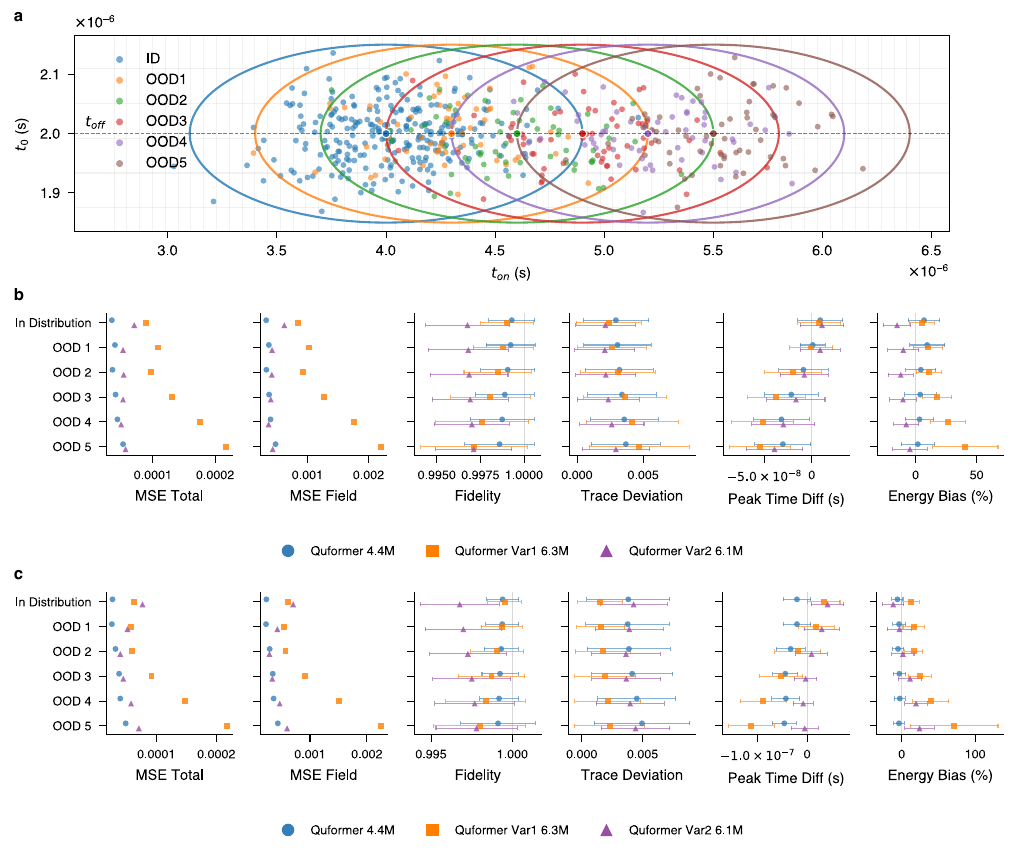}
\caption{\label{fig:quformer evaluation} \textbf{Model performance evaluation details.} \textbf{a}, Data-generating distribution for the control field turn-on time $t_{\text{on}}$ and probe pulse arrival time $t_0$, used to define in-distribution (ID) and out-of-distribution (OOD) test regimes. OOD datasets extend up to $5\sigma$ shifts from the training distribution mean in $t_{\text{on}}$. The solid ovals indicate the $3\sigma$ region of the Gaussian distribution, and each point corresponds to one sampled data instance.  \textbf{b}, Evaluation of model performance in the decoherence-free setting, where the spin wave evolves unitarily during storage. All three Quformer variants maintain high fidelity, low trace deviation, and low token-level mean squared error (MSE) across the full OOD range. Degradation in field observables (energy and peak time) emerges gradually and is most pronounced at extreme shifts (OOD 4–5) for Quformer Var1 (6.3M). Error bars represent one standard deviation over the test set. \textbf{c}, Evaluation under decoherence, where the quantum memory evolves as an open system. While quantum state fidelity remains robust ($>0.99$), output field metrics exhibit increased sensitivity to control timing shifts. At OOD 4–5, partial truncation of the retrieval pulse occurs in both prediction and target due to the finite simulation window. Across all settings, the Quformer (4.4M) consistently demonstrates the strongest performance across all evaluation axes. Error bars represent one standard deviation over the test set.}
\end{figure*}

We focus on two independent parameters in the data-generating process that most closely reflect the operating conditions of real-world quantum memory experiments. First, we fix the control field's turn-off time, \( t_{\text{off}}=2.0\ \mu s \). In experimental settings, the arrival time \( t_0 \) of the probe field pulse typically exhibits temporal jitter. To emulate this behavior, we sample \( t_0 \) from a Gaussian distribution centered around \( t_{\text{off}} \). Additionally, the turn-on time \( t_{\text{on}} \) of the control field determines the readout time of the stored spin excitation. Accordingly, we also draw \( t_{\text{on}} \) values from a Gaussian distribution. The complete data-generating distribution is summarized in Table~\ref{tab:data-generating-transposed} and illustrated in Fig.~\ref{fig:quformer evaluation}a.

We assess model performance using a comprehensive suite of evaluation metrics, grouped into three categories: model-level errors, tomography based metrics, and experimentally relevant observables. Model-level metrics quantify the neural network’s ability to reconstruct its direct outputs across the spatial and temporal lattice. These include the mean squared error at the token level (Token MSE), and the mean squared error between predicted and target electric field amplitudes (Field MSE). Tomography-based metrics evaluate the predicted quantum states trajectories themselves: we compute the average fidelity between predicted and ground-truth density matrices, along with the mean deviation of the trace from unity, which is a diagnostic of physical constraint violation. Finally, to connect the surrogate’s output to laboratory observables, we introduce two experiment-aligned metrics: the discrepancy in the timing of the photon readout peak (Peak Time Difference), and the relative error in the integrated pulse energy (Energy Bias). Together, these metrics provide a multifaceted evaluation of both quantum state reconstruction accuracy and experimental realism. A complete summary of definitions appears in Table~\ref{tab:evaluation_metrics}.

\begin{table}[htbp]
\centering
\caption{Evaluation metrics used to assess the model performance.}
\label{tab:evaluation_metrics}
\begin{tabular}{p{6.5cm} p{5.5cm}}
\hline
\multicolumn{2}{l}{\textit{Model-Level Metric}} \\
\hline
\parbox[t]{5.5cm}{\raggedright Token MSE ($\text{MSE}_{\text{token}}$)} & $\displaystyle \text{MSE}_{\text{token}} = \frac{1}{N} \sum_{i=1}^{N} \left\| \mathbf{z}_i^{\text{pred}} - \mathbf{z}_i^{\text{true}} \right\|^2$\\
\parbox[t]{5.5cm}{\raggedright E-field MSE ($\text{MSE}_E$)} & $\displaystyle \text{MSE}_E = \frac{1}{N} \sum_{i=1}^{N} \left| E_i^{\text{pred}} - E_i^{\text{true}} \right|^2$ \\
\hline
\multicolumn{2}{l}{\textit{Tomography-based Metrics}} \\
\hline
\parbox[t]{5.5cm}{\raggedright Avg. Fidelity ($\overline{F}$)} & $\overline{F} = \frac{1}{N} \sum_{i=1}^{N} \left( \text{Tr} \sqrt{ \sqrt{\rho_i^{\text{true}}} \, \rho_i^{\text{pred}} \, \sqrt{\rho_i^{\text{true}}} } \right)^2$ \\
\parbox[t]{5.5cm}{\raggedright Avg. Trace Deviation ($\overline{\Delta \text{Tr}}$)} & $\displaystyle \overline{\Delta \text{Tr}} = \frac{1}{N} \sum_{i=1}^{N} \left| \text{Tr}(\rho_i^{\text{pred}}) - 1 \right|$ \\
\hline
\multicolumn{2}{l}{\textit{Experimental Observation Metrics}} \\
\hline
\parbox[t]{5.5cm}{\raggedright Readout Time Difference ($\Delta t$)} & $\displaystyle \Delta t = \left| t_{\text{readout}}^{\text{pred}} - t_{\text{readout}}^{\text{true}} \right|$ \\
\parbox[t]{8cm}{\raggedright Energy Bias ($\Delta \eta / \eta_{\text{true}}$)} & $\displaystyle \Delta \eta = \frac{ \int E_{\text{pred}}^2(t)\,dt - \int E_{\text{true}}^2(t)\,dt }{ \int E_{\text{true}}^2(t)\,dt } $ \\
\hline
\end{tabular}
\end{table}

We begin by evaluating model performance in the decoherence-free setting, where the spin wave undergoes unitary evolution without decoherence during the storage interval. We consider a series of datasets defined by increasing shifts in the control field turn-on time $t_{on}$, ranging from in-distribution (ID) to $5\sigma$ out-of-distribution (OOD) deviations (Fig.~\ref{fig:quformer evaluation}a). We focus on evaluating a representative baseline and two principled model variants of the architecture. While broader exploration of model configurations may yield improved performance, such optimization lies beyond the scope of the present work. Across all Quformer variants, performance degrades gracefully with increasing OOD level, indicating robust inductive generalization rather than collapse. Importantly, the predicted quantum states maintain consistently fidelity $>0.995$ and low trace deviation across the entire shift range, demonstrating that the models preserve essential physical structure even under extrapolated timing conditions (Fig.~\ref{fig:quformer evaluation}b). The most prominent impact of the timing shift appears in the peak retrieval time difference and output energy bias. Among the three variants, the Quformer (4.4M) exhibits the smallest peak shift and lowest energy bias across all conditions, as well as the highest states fidelity, suggesting that moderate model capacity and balanced architecture promotes physically stable generalization. In contrast, Quformer Var1 (6.3M) undergoes a more pronounced increase in MSE at higher OOD levels, indicating reduced robustness to control-field variation. Overall, these results show that the Quformer 4.4M architecture most effectively captures the quantum memory dynamics and maintains physically meaningful generalization across a broad range of timing variability.

We next assess model robustness in the presence of decoherence, where the quantum memory spin wave undergoes dissipative evolution during storage(see Methods). Across all Quformer variants, the predicted quantum states retain high fidelity ($>0.99$) and low trace deviation(Fig.~\ref{fig:quformer evaluation}c), indicating that the surrogate accurately captures open-system dynamics even as unitary evolution is disrupted by loss and dephasing. Compared to the decoherence-free setting, output field metrics such as peak retrieval time and energy bias exhibit greater sensitivity to OOD timing shifts. This behavior may reflect the limited temporal coverage of the training data, which could restrict the model’s ability to fully learn the decoherence rate from observed decay trajectories. At the most extreme shifts (OOD 4–5), partial truncation of the output pulse occurs in both the predicted and reference trajectories due to the finite simulation window. While this may influence energy metrics in some cases, all models continue to preserve internal state fidelity and maintain physically plausible predictions. Among the three variants, the Quformer (4.4M) remains the most stable across all evaluation metrics, while Quformer Var1 shows a sharper increase in MSE and output timing drift. These results reinforce the model’s ability to generalize not only across extrapolated control parameters, but also under realistic decoherence dynamics, validating its applicability to open quantum systems.

\begin{figure*}
\includegraphics[width=\linewidth]{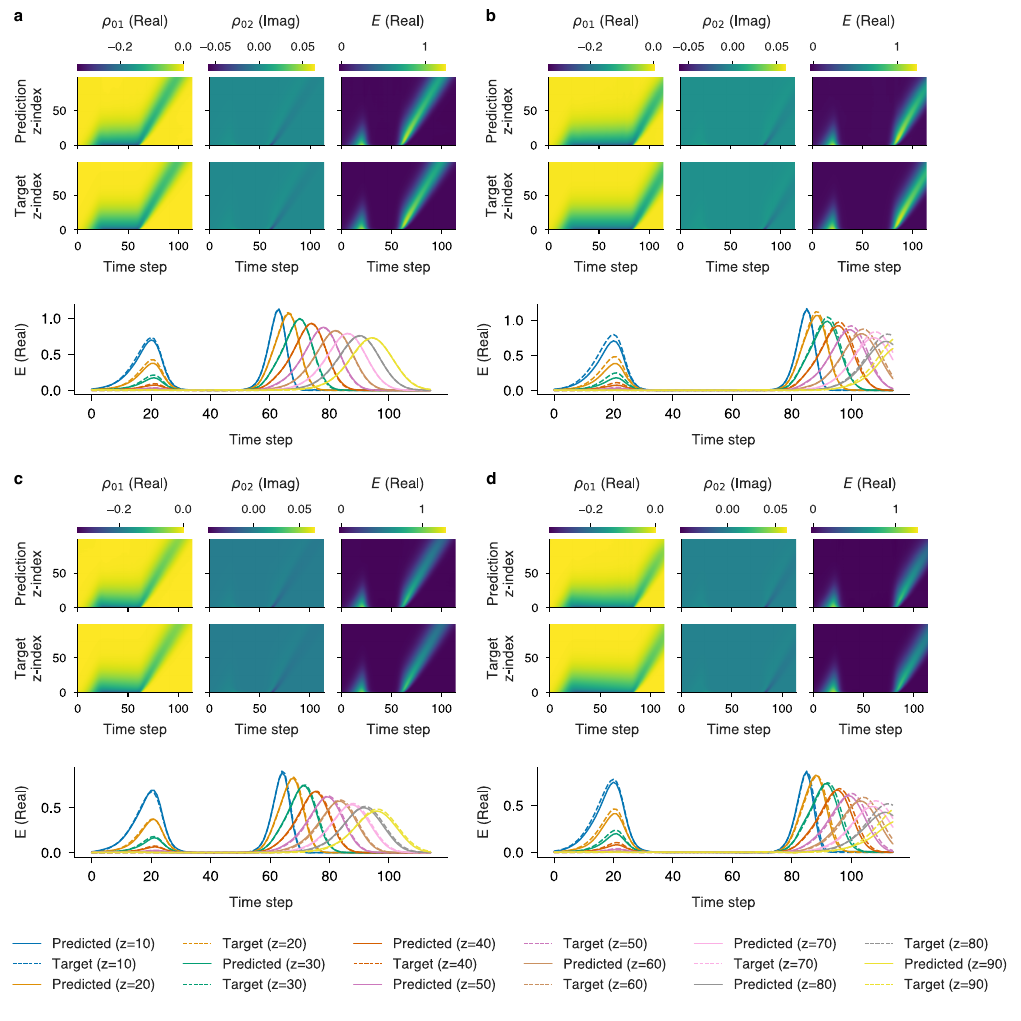}
\caption{\label{fig:prediction_target} \textbf{Predicted trajectories examples for key observables.} Predicted (top row) and target (bottom row) spatiotemporal trajectories are shown for three physically significant quantities: the spin coherence  $\rho_{01}$ (real part), the polarization density corresponding $\rho_{02}$ (imaginary part), and the probe field envelope $E$ (real part). Each panel displays a distinct test condition: \textbf{a}, decoherence-free in-distribution; \textbf{b}, decoherence-free at $5\sigma$ out-of-distribution (OOD); \textbf{c}, decoherence in-distribution; and \textbf{d}, decoherence at $5\sigma$ OOD. The heatmaps show the observable value across space (vertical axis) and time (horizontal axis). Line plots below show the real part of the experimental observable $E$ as a function of time for 9 equally spaced spatial positions, comparing predicted (solid) and ground-truth (dashed) trajectories. Despite strong extrapolation in both control parameters and system dynamics, the Quformer (4.4M) model generates smooth, well-aligned predictions without spurious oscillations or collapse.}
\end{figure*}

We further examine representative spatiotemporal trajectories of three physically significant observables: the spin coherence $\rho_{01}$, the polarization density $\rho_{02}$, and the probe field envelope $E$, visualized across space and time (Fig.~\ref{fig:prediction_target}). These quantities are closely tied to the memory's dynamics and experimental observables: $\rho_{01}$ encodes the spin-wave amplitude, $\rho_{02}$ governs the light--matter interaction strength via polarization, and $E$ corresponds to the directly measurable probe field. In the in-distribution setting, the model accurately reproduces both spatial and temporal features of the target trajectories across all three observables. At the $5\sigma$ out-of-distribution shift where the control field activation lies far outside the training region, the predictions remain smooth, physically consistent, and aligned in structure with the ground truth. Temporal shifts are visible in the retrieved probe pulse envelope at $5\sigma$ OOD sets(Fig.~\ref{fig:prediction_target}b, d), consistent with earlier quantitative metrics (Fig.~\ref{fig:quformer evaluation}b--c). The model does not exhibit collapse or spurious oscillations. These results provide confirmation that the model preserves accurate spatiotemporal structure and faithfully reproduces the system’s underlying physical dynamics even under nontrivial extrapolation and open-system decoherence.

\begin{figure*}[htbp]
\includegraphics[width=\linewidth]{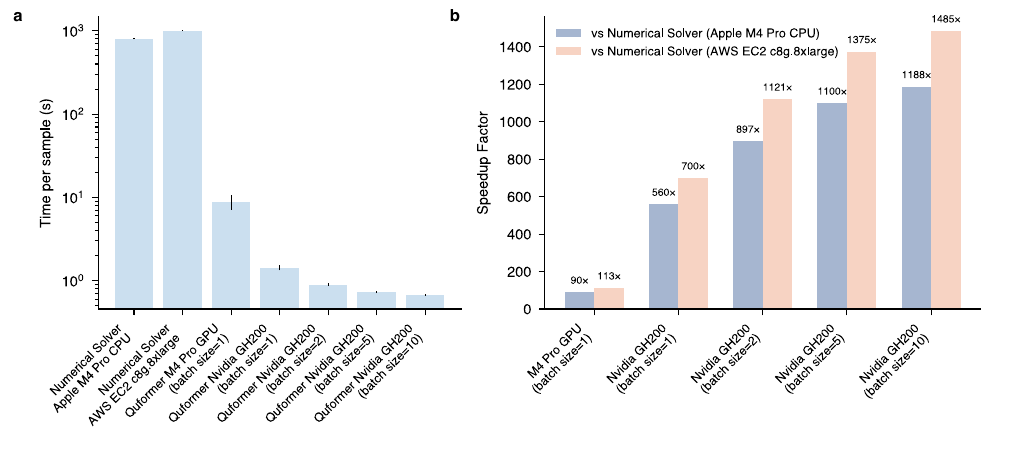}
\caption{\label{fig:quformer acceleration} \textbf{Acceleration of the deep learning model over numerical solver.} \textbf{a}, Runtime per sample for the numerical solver and the deep learning model, evaluated on different hardware platforms and batch sizes. The solver is executed on both an Apple M4 Pro CPU (MacBook Pro) and an AWS EC2 c8g.8xlarge instance, while inference is performed on the Apple GPU (via PyTorch MPS) and a cloud-based NVIDIA GH200 GPU. \textbf{b}, Corresponding speedup factors of the surrogate relative to the classical solver. Blue bars indicate acceleration over the M4 Pro CPU baseline; orange bars indicate acceleration over the AWS CPU baseline. These results provide a representative lab-scale benchmark, demonstrating the surrogate model's ability to achieve substantial computational gains (up to $\sim$1485$\times$) even without backend-specific optimization. }
\end{figure*}

To evaluate the practical computational advantage of the Quformer model on quantum memory learning and serve as a representative lab-scale reference to illustrate acceleration potential, we benchmark its inference time against the classical numerical solver on which we generate the training data. Both implementations were written in Python using standard libraries, and no explicit low-level optimization was applied. On an Apple M4 Pro chip, we observe a 90× acceleration when running the model inference on the Apple GPU (via PyTorch’s MPS backend), compared to solving the Maxwell–Bloch-type equations on its CPU. We also report a 113× acceleration relative to an AWS EC2 c8g.8xlarge CPU instance. To assess scalability, we deploy the trained model on a cloud-based NVIDIA GH200 GPU, where inference achieves speedup factors ranging from 560× to 1485×, depending on batch size (1 to 10). These measurements reflect runtime per sample and are summarized in Fig.~\ref{fig:quformer acceleration}. Notably, the classical solver requires adaptive time stepping to maintain numerical stability, typically producing up to $10^5$ time points per simulation. In contrast, the Quformer model directly predicts observables on a uniform 120-point time grid. This difference reflects not only architectural speedups, but also the model’s ability to bypass numerical stiffness that constrains the numerical solver. While neither numerical solver nor deep learning model was manually tuned for peak speed, the results demonstrate that even modest batch inference on modern hardware can achieve orders-of-magnitude reduction in simulation time. 

\section{Discussion}

In summary, we introduced a regional-attention–based neural architecture for modeling the control-driven dynamics of spatially structured open quantum systems. By embedding physics-informed inductive biases, conditioning on time-dependent fields, and combining scalable local–global attention mechanism, the model achieves high predictive fidelity across representative testbeds while reducing simulation time by up to three orders of magnitude relative to the classical solver. These results establish the framework as a general-purpose surrogate for structured open quantum system dynamics, bridging accuracy, scalability, and efficiency.

The acceleration and fidelity achieved here open up versatile potential applications in quantum information science. Surrogate models of this type could enhance large-scale quantum network simulations\cite{coopmans2021netsquid, bartlett2018distributed, wu2021sequence} by providing physically grounded, time-dependent device dynamics, enabling near real-time exploration of protocols, scheduling strategies, and throughput analysis. They provide a tool for repeater and memory design, where accurate modeling of timing-dependent control pulses is essential for multiplexing and asynchronous operation, as well as timing-dependent success rates. In laboratory settings, fast surrogates could be embedded into adaptive feedback loops for pulse shaping, Bayesian parameter optimization, or control scheduling, accelerating experimental progress. The framework also generalizes naturally to scalable device modeling, with potential relevance to cavity QED, waveguide QED, and atomic array platforms where spatiotemporal structure and time dependent driving govern performance.

Beyond simulation-driven applications, an important future direction of our work is the integration of experimental data. Real-world devices often exhibit imperfections and noise processes not captured by idealized theory models. An example is the four-wave-mixing induced noise in EIT quantum memory. Training the surrogate directly on experimental datasets, or jointly on hybrid simulation–experiment datasets, could enable noise-aware modeling of device behavior. Such models could then support rapid parameter searches, adaptive calibration, and predictive optimization of quantum memories and related devices. Thus, the framework can serve as a practical experimental tool, accelerating refinement of quantum-device performance.

The present framework has two primary limitations that define its scope. First, long-range entanglement between distant regions is not explicitly represented, since each grid site is modeled by its local reduced density matrix token. Extensions that incorporate small multi-site clusters may capture short-range entanglement beyond single-site states. Second, model performance is bounded by the coverage of its training data. While we demonstrated robust generalization across control-field variations, extrapolation to untrained regimes is naturally constrained. Incorporating broader datasets, especially experimental data as noted above, provides a clear path to mitigate this limitation.

In conclusion, this work establishes a physics-informed, scalable regional-attention architecture for surrogate modeling of structured open quantum dynamics. By balancing fidelity, speed, and extensibility, the framework opens a path from foundational modeling of light–matter interfaces to practical engineering of quantum networks, repeaters, and device optimization in realistic experimental settings.

\section{Methods}

\subsection{Synthetic data generation for single qubit learning}

We generated trajectories of a driven, dissipative single-qubit system using the QuTiP simulation library\cite{lambert2024qutip5quantumtoolbox}. The control field follows a sigmoidal temporal profile parameterized by a turn-off time $t_{\text{off}}$, which sets the duration of coherent driving. The time axis is defined in arbitrary units, with one unit representing the full evolution window. The driving strength is fixed to produce approximately three Rabi oscillations over the unit interval. The amplitude damping rate is set such that the excited-state population decays to 10\% of its initial value by the end of the trajectory. Each trajectory consists of $T = 240$ time steps with uniform spacing $\Delta t = 1/240$, and is initialized from a pure quantum state sampled from the Haar measure on the Bloch sphere. The datasets are generated by sampling $t_{\text{off}}$ from Gaussian distributions: the in-distribution (ID) set uses a mean of 0.5 with standard deviation 0.05, while out-of-distribution (OOD) sets OOD1 to OOD3 use means of 0.55, 0.60, and 0.65, respectively. The full dataset includes 700 training trajectories, 150 validation trajectories, and 100 test trajectories for each of the ID and OOD conditions.

\subsection{Synthetic data generation for quantum memory learning}

The training data for the quantum memory learning were generated using high-fidelity simulations of a $\Lambda$-type EIT quantum memory in a rubidium-87 atomic ensemble. The system consists of three energy levels: ground states $\ket{0} = \ket{5S_{1/2}, F=1, m=0}$ and $\ket{1} = \ket{5S_{1/2}, F=2, m=0}$, and an excited state $\ket{2} = \ket{5P_{1/2}, F=1, m=-1}$, with a 6.8\,GHz hyperfine splitting between the ground states. The ensemble population is initialized in a thermal distribution at $100\,^\circ\mathrm{C}$, with an atomic density of $4 \times 10^{17}\,\text{atoms}/\text{m}^3$.

Two laser fields drive the transitions: a weak, Gaussian-shaped probe field resonant with the $\ket{0} \leftrightarrow \ket{2}$ transition and a strong, sigmoidal control field coupling the $\ket{1} \leftrightarrow \ket{2}$ transition. Field amplitudes are derived from experimentally relevant parameters: 0.4\,mW control field, 0.01\,mW probe field, and a 1.6\,mm beam diameter.

To investigate different coherence regimes, we simulate two classes of datasets. The decoherence-free dataset includes only spontaneous emission from the excited state at a rate of $2\pi \times 5.746$\,MHz. The decoherence-included dataset models a buffer-gas-filled, anti-relaxation-coated vapor cell, incorporating two additional ground-state decoherence channels: a dephasing rate of $2\pi \times 2.5$\,kHz and a population decay rate of $2\pi \times 2.5$\,kHz during the storage phase. These channels account for spinwave decoherence due to atomic collisions and motional effects.

System dynamics are governed by the coupled equations defined in Eqs.~(7) and (8). The quantum master equation is solved using an exponential time differencing scheme, while the propagation of electromagnetic fields is computed using a fourth-order Adams–Bashforth–Moulton method. Spatial discretization spans 100 points over a 1\,cm medium. Adaptive time stepping ensures numerical stability by enforcing a 1\% threshold on variations in both the density matrix and boundary field values between successive steps.

Training and test data are generated by sampling experimental parameters according to the data-generating process described in Table~\ref{tab:data-generating-transposed}. Each simulation covers a total evolution time of $7.5\,\mu\text{s}$. Solutions, initially computed on non-uniform time grids due to adaptive stepping, are interpolated onto a uniform 128-point temporal grid. For training purposes, we select a spatial–temporal subgrid of dimension 120 (time) $\times$ 99 (space) that captures the complete storage and readout process as the model’s region of interest.

\begin{table}[ht]
\centering
\scriptsize
\setlength{\tabcolsep}{3pt}
\renewcommand{\arraystretch}{1.0}
\caption{\textbf{Statistical summary of data generating process.}
Each parameter is sampled from a Gaussian distribution. Means and standard deviations ($\sigma$) are shown for in-distribution (ID) and out-of-distribution (OOD) datasets.}
\label{tab:data-generating-transposed}
\begin{tabular}{@{}lcccccc@{}}
\toprule
\textbf{Attribute} & \textbf{ID} & \textbf{OOD 1} & \textbf{OOD 2} & \textbf{OOD 3} & \textbf{OOD 4} & \textbf{OOD 5} \\
 & & ($+1\sigma$) & ($+2\sigma$) & ($+3\sigma$) & ($+4\sigma$) & ($+5\sigma$) \\
\midrule
\# Samples & 260 & 60 & 60 & 60 & 60 & 60 \\
Purpose & Train/Val/Test & Test & Test & Test & Test & Test \\
\midrule
$t_0$ Mean ($\mu$s) & 2.00 & 2.00 & 2.00 & 2.00 & 2.00 & 2.00 \\
$t_0$ Std. ($10^{-2}\,\mu$s) & 5.00 & 5.00 & 5.00 & 5.00 & 5.00 & 5.00 \\
\midrule
$t_{\text{on}}$ Mean ($\mu$s) & 4.00 & 4.30 & 4.60 & 4.90 & 5.20 & 5.50 \\
$t_{\text{on}}$ Std. ($10^{-1}\,\mu$s) & 3.00 & 3.00 & 3.00 & 3.00 & 3.00 & 3.00 \\
\bottomrule
\end{tabular}
\end{table}

\subsection{Model variation details}

Three model configurations were evaluated in quantum memory learning: Quformer 4.4M (baseline), Quformer Var1 6.3M, and Quformer Var2 6.1M. All variants used a token embedding dimension of 11 and 4 attention heads per layer. Quformer 4.4M contains approximately 4.4 million trainable parameters and comprises $N_B = 4$ attention blocks. Each block includes two regional self-attention layers with axial type decomposition: the first layer applies attention along the temporal axis with local region of shape $(T, Z)=(120, 1)$, and the second along the spatial axis with shape $(1, 99)$. Quformer Var1 6.3M uses the same axial decomposition per block but increases the number of attention blocks to $N_B = 6$, resulting in approximately 6.3 million parameters. Quformer Var2 6.1M retains $N_B = 4$ attention blocks but employs a three-stage regional decomposition within each block: the first layer attends over $(120, 1)$, followed by a layer with region shape $(12, 9)$, and a third layer with shape $(1, 99)$. The total number of trainable parameters is approximately 6.1 million. A table summarize the variants difference is in supplemental material.

\subsection{Training}
The Quformer model for quantum memory learning is trained using the root-mean-square error (RMSE) loss: $\mathcal{L}_{\text{total}}(\boldsymbol{\theta}) = \sqrt{\frac{1}{N} \sum_{i=1}^{N}\left(X_{\textbf{predict}, i} - X_{\textbf{target}, i}\right)^2}$. Optimization is performed using the AdamW algorithm with a weight decay coefficient of 0.1, applied only to weight matrix parameters. The model is trained on 140 unique in-distribution examples with a validation set of 30 examples. Training is performed for 150 epochs using a batch size of 5. The learning rate follows a schedule with linear warm-up over the first 150 steps to a peak of $1.0\times10^{-3}$, followed by cosine decay to a minimum of $1.0\times10^{-6}$. All training is conducted on an NVIDIA GH200 GPU.

The single-qubit model is trained using the same RMSE loss function, AdamW optimizer, and weight decay configuration as the quantum memory model. The learning rate schedule consists of a linear warm-up over the first 150 steps to a peak of $1.1\times10^{-3}$, followed by cosine decay to a minimum of $1.0\times10^{-7}$. The model is trained on 700 in-distribution examples and validated on 150 examples over 150 epochs with a batch size of 10.

\subsection{Metrics evaluation}
In the quantum memory learning model performance was evaluated using 60 test examples per OOD region, except for OOD4 in the decoherence-free and decoherence learning settings, which included 59 examples. All metrics were computed over the full test set within each OOD region. Token mean squared error (token MSE) and electric field mean squared error (E-field MSE) were calculated by aggregating the squared errors over all tokens across all test examples, followed by averaging over the entire set. Average trace deviation was computed as the absolute deviation of the predicted density matrix trace from unity at each spacetime grid point, averaged over all tokens in the test set. Predicted density matrices were then normalized to have unit trace, and fidelity with respect to the ground-truth density matrices was computed at each token and averaged over the full set. The readout time difference was defined as the temporal offset between the predicted and ground-truth peaks of the retrieved probe pulse. For each example, the readout time was identified as the temporal location of the global maximum of the probe field amplitude, and the difference was averaged over all test examples in the OOD region. Energy bias was computed by integrating the probe field intensity at the spatial grid position $z=64$ (corresponding to z=0.64cm along the rubidium cell), and averaging over all examples. This position ensured full pulse capture within the simulation time window for test sets up to OOD3, while pulses were partially truncated for OOD4 and OOD5.

For the single-qubit learning, model performance was evaluated using 100 test examples per OOD region. The token MSE, average fidelity, and average trace deviation were computed following the same procedures as in the quantum memory evaluation, by aggregating metrics over all tokens from all test examples within each OOD region.

\subsection{Acceleration evaluation}

To quantify the computational performance of the surrogate model relative to classical numerical integration, we benchmarked both methods on identical simulation tasks. We use a total of 10 examples from the in-distribution test dataset. For the classical numerical solver, each example was executed once and the reported timing is the average over all 10 runs. Inference time for the Quformer model was evaluated using batch sizes of 1, 2, 5, and 10. For batch sizes greater than 1, the number of runs was adjusted such that a total of 10 batches were processed. The per-sample inference time was estimated by dividing the total batch runtime by the batch size. On the Apple M4 Pro MacBook Pro, inference was performed using the PyTorch Metal Performance Shaders (MPS) backend to leverage the integrated Apple GPU. For high-performance inference benchmarking, the model was deployed on a Lambda Cloud instance equipped with an NVIDIA GH200 GPU using the PyTorch CUDA backend.

The classical solver uses an adaptive time-stepping scheme and typically generates approximately $10^5$ uneven time steps per trajectory to maintain numerical stability. In contrast, the surrogate model operates on a fixed 120-point uniform time grid, corresponding to linearly interpolated outputs from the solver, and this grid was used consistently during training and evaluation. Direct numerical integration on such a coarse grid is unstable and results in divergent or non-physical trajectories, rendering the output unusable for downstream analysis. Accordingly, all runtime comparisons were performed between the solver’s adaptive time stepping trajectories and the model’s predictions on the fixed grid.

\backmatter

\bmhead{Supplementary information} 

Supplementary information is in the Appendix sections. 

\bmhead{Data availability} 

All data for reproduce this work and evaluations are available in the main text or the
supplementary materials.

\bmhead{Code availability} 

The codebase and trained model can be found at \url{https://github.com/Dounan662/RegionalAttentionOpenQuantum}

\bmhead{Acknowledgements}
This work was supported by the Stony Brook Foundation (Quantum Network Research Center) and by the U.S. National Science Foundation through the National Quantum Virtual Laboratory (NQVL) QSTD Pilot project ``SCY-QNet: A Wide-Area Quantum Network to Demonstrate Quantum Advantage'' (Award No.\ 2410725).

\bmhead{Competing interests} 
The authors declare no competing interests.

\bmhead{Author contribution} 
D.D. conceived the problem and developed the algorithms, conducted the analysis and organized the manuscript. E.F. supervised the project. All authors discussed the results and contributed to the final manuscript.










\begin{appendices}

\section{Model variants details}\label{secA1}

\begin{table}[h]
\centering
\setlength{\tabcolsep}{4pt}
\caption{\textbf{Summary of Quformer model variants.} Each variant differs in the number of attention blocks and the regional decomposition strategy. All models use 4-headed attention and a token embedding dimension of 11.}
\label{tab:quformer_variants}
\begin{tabular}{lcccc}
\hline
\textbf{Model} & \textbf{Parameters} & \boldmath$N_B$ & \textbf{Regional Decomposition} & \textbf{Layers per Block} \\
\hline
Quformer 4.4M (baseline) & 4.4M & 4 & $(120,1) \rightarrow (1,99)$ & 2 \\
Quformer Var1 6.3M & 6.3M & 6 & $(120,1) \rightarrow (1,99)$ & 2 \\
Quformer Var2 6.1M & 6.1M & 4 & $(120,1) \rightarrow (12,9) \rightarrow (1,99)$ & 3 \\
\hline
\end{tabular}
\end{table}

\section{Model Evaluation Details}\label{secA2}

\begin{threeparttable}
\centering
\small
\caption{\textbf{Model performance on in-distribution (ID) and out-of-distribution (OOD) datasets in decoherence free regime.} 
Best per-row values are bolded.}
\label{tab:perf}

\renewcommand{\arraystretch}{0.5}
\small
\begin{tabular}{l|@{\hspace{0.3em}}l @{\hspace{0.3em}}|@{\hspace{0.3em}} p{2.3cm} p{2.3cm} p{2.3cm}}
\toprule
 & & \multicolumn{3}{c}{\textbf{Model}} \\
\cmidrule(lr){3-5}
\textbf{Dataset} & \textbf{Metric}\tnote{a} 
  & \text{Quformer(4.4M)} & \text{Var1(6.3M)} & \text{Var2(6.1M)} \\
\midrule
\multirow{9}{*}{ID} 
  & $\text{MSE}_{\text{token}}$
    & {\bfseries 0.000034} & 
0.000089 & 
0.000070 \\
  & $\text{MSE}_E$
    & {\bfseries 0.000324} & 
0.000838 & 
0.000621 \\
  & $\overline{F}$ 
    & {\bfseries 0.999297} & 
0.999024 & 
0.996758 \\
  & $\sigma_F$
    & {\bfseries 0.001329} & 
0.001512 & 
0.002402 \\
  & $\overline{\Delta \text{Tr}}$ 
    & 0.002939 & 
0.002384 & 
{\bfseries 0.002144} \\
  & $\sigma_{\Delta \text{Tr}}$
    & 0.002484 & 
0.002453 & 
{\bfseries 0.002252} \\
  & $\Delta t$(ns)
    & $8.8\pm23.5$ & 
$\textbf{7.8}\pm\textbf{22.6}$ & 
$10.7\pm22.7$ \\
  & $\Delta \eta / \eta_{\text{true}}$(\%)
    & $6.89\pm12.78$ & 
$\textbf{5.01}\pm\textbf{10.32}$ & 
$-15.22\pm11.11$ \\
\midrule
\multirow{9}{*}{OOD1} 
  & $\text{MSE}_{\text{token}}$
    & {\bfseries 0.000039} & 
0.000108 & 
0.000052 \\
  & $\text{MSE}_E$
    & {\bfseries 0.000368} & 
0.001031 & 
0.000421 \\
  & $\overline{F}$ 
    & {\bfseries 0.999231} & 
0.998811 & 
0.996808 \\
  & $\sigma_F$
    & {\bfseries 0.001400} & 
0.001723 & 
0.002301 \\
  & $\overline{\Delta \text{Tr}}$ 
    & 0.003060 & 
0.002651 & 
{\bfseries 0.002102} \\
  & $\sigma_{\Delta \text{Tr}}$
    & 0.002535 & 
0.002595 & 
{\bfseries 0.002268} \\
  & $\Delta t$(ns)
    & $\textbf{1.0}\pm\textbf{13.1}$ & 
$-1.0\pm22.7$ & 
$8.8\pm20.9$ \\
  & $\Delta \eta / \eta_{\text{true}}$(\%)
    & $\textbf{9.48}\pm\textbf{14.22}$ & 
$10.24\pm12.01$ & 
$-10.11\pm12.76$ \\
\midrule
\multirow{9}{*}{OOD2} 
  & $\text{MSE}_{\text{token}}$
    & {\bfseries 0.000035} & 
0.000097 & 
0.000053 \\
  & $\text{MSE}_E$
    & {\bfseries 0.000328} & 
0.000925 & 
0.000416 \\
  & $\overline{F}$ 
    & {\bfseries 0.999064} & 
0.998493 & 
0.996856 \\
  & $\sigma_F$
    & {\bfseries 0.001539} & 
0.001934 & 
0.002226 \\
  & $\overline{\Delta \text{Tr}}$ 
    & 0.003205 & 
0.003090 & 
{\bfseries 0.002176} \\
  & $\sigma_{\Delta \text{Tr}}$
    & 0.002565 & 
0.002826 & 
{\bfseries 0.002275} \\
  & $\Delta t$(ns)
    & $-8.8\pm25.8$ & 
$-19.5\pm29.6$ & 
$\textbf{-7.8}\pm\textbf{25.0}$ \\
  & $\Delta \eta / \eta_{\text{true}}$(\%)
    & $\textbf{4.26}\pm\textbf{12.25}$ & 
$10.90\pm10.22$ & 
$-12.19\pm10.25$ \\
\midrule
\multirow{9}{*}{OOD3} 
  & $\text{MSE}_{\text{token}}$
    & {\bfseries 0.000040} & 
0.000130 & 
0.000052 \\
  & $\text{MSE}_E$
    & {\bfseries 0.000372} & 
0.001274 & 
0.000400 \\
  & $\overline{F}$ 
    & {\bfseries 0.998898} & 
0.998072 & 
0.996918 \\
  & $\sigma_F$
    & {\bfseries 0.001687} & 
0.002282 & 
0.002169 \\
  & $\overline{\Delta \text{Tr}}$ 
    & 0.003398 & 
0.003621 & 
{\bfseries 0.002358} \\
  & $\sigma_{\Delta \text{Tr}}$
    & 0.002588 & 
0.003128 & 
{\bfseries 0.002334} \\
  & $\Delta t$(ns)
    & $-21.5\pm28.2$ & 
$-37.1\pm30.2$ & 
$\textbf{-16.6}\pm\textbf{30.4}$ \\
  & $\Delta \eta / \eta_{\text{true}}$(\%)
    & $\textbf{3.72}\pm\textbf{13.02}$ & 
$17.49\pm12.23$ & 
$-10.32\pm11.01$ \\
\midrule
\multirow{9}{*}{OOD4} 
  & $\text{MSE}_{\text{token}}$
    & {\bfseries 0.000043} & 
0.000176 & 
0.000049 \\
  & $\text{MSE}_E$
    & 0.000395 & 
0.001759 & 
{\bfseries 0.000365} \\
  & $\overline{F}$ 
    & {\bfseries 0.998749} & 
0.997604 & 
0.997009 \\
  & $\sigma_F$
    & {\bfseries 0.001842} & 
0.002670 & 
0.002153 \\
  & $\overline{\Delta \text{Tr}}$ 
    & 0.003569 & 
0.004171 & 
{\bfseries 0.002628} \\
  & $\sigma_{\Delta \text{Tr}}$
    & 0.002584 & 
0.003464 & 
{\bfseries 0.002434} \\
  & $\Delta t$(ns)
    & $-31.8\pm29.2$ & 
$-51.6\pm32.5$ & 
$\textbf{-29.8}\pm\textbf{33.0}$ \\
  & $\Delta \eta / \eta_{\text{true}}$(\%)
    & $\textbf{3.21}\pm\textbf{11.76}$ & 
$26.55\pm14.49$ & 
$-7.69\pm10.37$ \\
\midrule
\multirow{9}{*}{OOD5} 
  & $\text{MSE}_{\text{token}}$
    & {\bfseries 0.000052} & 
0.000218 & 
0.000056 \\
  & $\text{MSE}_E$
    & 0.000478 & 
0.002199 & 
{\bfseries 0.000431} \\
  & $\overline{F}$ 
    & {\bfseries 0.998600} & 
0.997124 & 
0.997115 \\
  & $\sigma_F$
    & {\bfseries 0.001999} & 
0.003060 & 
0.002182 \\
  & $\overline{\Delta \text{Tr}}$ 
    & 0.003690 & 
0.004716 & 
{\bfseries 0.002940} \\
  & $\sigma_{\Delta \text{Tr}}$
    & 0.002605 & 
0.003774 & 
{\bfseries 0.002552} \\
  & $\Delta t$(ns)
    & $\textbf{-30.3}\pm\textbf{29.3}$ & 
$-54.7\pm31.9$ & 
$-39.1\pm29.6$ \\
  & $\Delta \eta / \eta_{\text{true}}$(\%)
    & $\textbf{1.81}\pm\textbf{13.72}$ & 
$40.61\pm26.88$ & 
$-4.65\pm14.07$ \\
\bottomrule
\end{tabular}
\begin{tablenotes}
\item[a] See Table ~\ref{tab:evaluation_metrics} for metric definition.  
\end{tablenotes}
\end{threeparttable}

\begin{threeparttable}
\centering
\small
\caption{\textbf{Model performance on in-distribution (ID) and out-of-distribution (OOD) datasets in decoherence regime.} 
Best per-row values are bolded.}
\label{tab:perf_deco}

\renewcommand{\arraystretch}{0.5}
\small
\begin{tabular}{l|@{\hspace{0.3em}}l @{\hspace{0.3em}}|@{\hspace{0.3em}} p{2.3cm} p{2.3cm} p{2.3cm}}
\toprule
 & & \multicolumn{3}{c}{\textbf{Model}} \\
\cmidrule(lr){3-5}
\textbf{Dataset} & \textbf{Metric}\tnote{a} 
  & \text{Quformer(4.4M)} & \text{Var1(6.3M)} & \text{Var2(6.1M)} \\
\midrule
\multirow{9}{*}{ID} 
  & $\text{MSE}_{\text{token}}$
    & {\bfseries 0.000027} & 
0.000063 & 
0.000077 \\
  & $\text{MSE}_E$
    & {\bfseries 0.000234} & 
0.000612 & 
0.000704 \\
  & $\overline{F}$ 
    & 0.999400 & 
{\bfseries 0.999519} & 
0.996761 \\
  & $\sigma_F$
    & {\bfseries 0.000984} & 
0.001024 & 
0.002427 \\
  & $\overline{\Delta \text{Tr}}$ 
    & 0.003815 & 
{\bfseries 0.001522} & 
0.004259 \\
  & $\sigma_{\Delta \text{Tr}}$
    & 0.003408 & 
{\bfseries 0.001844} & 
0.002749 \\
  & $\Delta t$(ns)
    & $\textbf{-20.5}\pm\textbf{27.9}$ & 
$33.2\pm32.7$ & 
$41.0\pm32.6$ \\
  & $\Delta \eta / \eta_{\text{true}}$(\%)
    & $\textbf{-5.90}\pm\textbf{7.60}$ & 
$12.01\pm12.00$ & 
$-11.66\pm14.27$ \\
\midrule
\multirow{9}{*}{OOD1} 
  & $\text{MSE}_{\text{token}}$
    & {\bfseries 0.000026} & 
0.000058 & 
0.000052 \\
  & $\text{MSE}_E$
    & {\bfseries 0.000228} & 
0.000551 & 
0.000427 \\
  & $\overline{F}$ 
    & {\bfseries 0.999374} & 
0.999342 & 
0.996982 \\
  & $\sigma_F$
    & {\bfseries 0.001004} & 
0.001269 & 
0.002352 \\
  & $\overline{\Delta \text{Tr}}$ 
    & 0.003766 & 
{\bfseries 0.001556} & 
0.003896 \\
  & $\sigma_{\Delta \text{Tr}}$
    & 0.003431 & 
{\bfseries 0.001970} & 
0.002790 \\
  & $\Delta t$(ns)
    & $-20.5\pm27.9$ & 
$\textbf{18.6}\pm\textbf{36.3}$ & 
$29.3\pm34.7$ \\
  & $\Delta \eta / \eta_{\text{true}}$(\%)
    & $-3.66\pm8.91$ & 
$16.16\pm14.50$ & 
$\textbf{-3.03}\pm\textbf{16.94}$ \\
\midrule
\multirow{9}{*}{OOD2} 
  & $\text{MSE}_{\text{token}}$
    & {\bfseries 0.000032} & 
0.000060 & 
0.000040 \\
  & $\text{MSE}_E$
    & 0.000292 & 
0.000570 & 
{\bfseries 0.000288} \\
  & $\overline{F}$ 
    & {\bfseries 0.999312} & 
0.999067 & 
0.997263 \\
  & $\sigma_F$
    & {\bfseries 0.001036} & 
0.001615 & 
0.002345 \\
  & $\overline{\Delta \text{Tr}}$ 
    & 0.003872 & 
{\bfseries 0.001719} & 
0.003632 \\
  & $\sigma_{\Delta \text{Tr}}$
    & 0.003416 & 
{\bfseries 0.002268} & 
0.002829 \\
  & $\Delta t$(ns)
    & $-33.2\pm29.0$ & 
$-18.6\pm46.0$ & 
$\textbf{8.8}\pm\textbf{31.8}$ \\
  & $\Delta \eta / \eta_{\text{true}}$(\%)
    & $-5.05\pm7.73$ & 
$16.13\pm12.33$ & 
$\textbf{1.54}\pm\textbf{15.12}$ \\
\midrule
\multirow{9}{*}{OOD3} 
  & $\text{MSE}_{\text{token}}$
    & {\bfseries 0.000038} & 
0.000092 & 
0.000045 \\
  & $\text{MSE}_E$
    & 0.000346 & 
0.000916 & 
{\bfseries 0.000339} \\
  & $\overline{F}$ 
    & {\bfseries 0.999241} & 
0.998715 & 
0.997506 \\
  & $\sigma_F$
    & {\bfseries 0.001111} & 
0.002032 & 
0.002391 \\
  & $\overline{\Delta \text{Tr}}$ 
    & 0.004138 & 
{\bfseries 0.001927} & 
0.003652 \\
  & $\sigma_{\Delta \text{Tr}}$
    & 0.003325 & 
{\bfseries 0.002522} & 
0.002832 \\
  & $\Delta t$(ns)
    & $-43.9\pm25.4$ & 
$-51.8\pm42.9$ & 
$\textbf{-2.9}\pm\textbf{22.5}$ \\
  & $\Delta \eta / \eta_{\text{true}}$(\%)
    & $\textbf{-3.26}\pm\textbf{8.45}$ & 
$24.47\pm15.45$ & 
$11.31\pm15.86$ \\
\midrule
\multirow{9}{*}{OOD4} 
  & $\text{MSE}_{\text{token}}$
    & {\bfseries 0.000040} & 
0.000148 & 
0.000058 \\
  & $\text{MSE}_E$
    & {\bfseries 0.000361} & 
0.001509 & 
0.000469 \\
  & $\overline{F}$ 
    & {\bfseries 0.999181} & 
0.998349 & 
0.997682 \\
  & $\sigma_F$
    & {\bfseries 0.001200} & 
0.002451 & 
0.002445 \\
  & $\overline{\Delta \text{Tr}}$ 
    & 0.004520 & 
{\bfseries 0.002116} & 
0.003974 \\
  & $\sigma_{\Delta \text{Tr}}$
    & 0.003195 & 
{\bfseries 0.002710} & 
0.002802 \\
  & $\Delta t$(ns)
    & $-42.7\pm28.2$ & 
$-88.4\pm46.2$ & 
$\textbf{-7.9}\pm\textbf{20.1}$ \\
  & $\Delta \eta / \eta_{\text{true}}$(\%)
    & $\textbf{-2.74}\pm\textbf{7.28}$ & 
$39.55\pm24.40$ & 
$19.47\pm15.19$ \\
\midrule
\multirow{9}{*}{OOD5} 
  & $\text{MSE}_{\text{token}}$
    & {\bfseries 0.000049} & 
0.000218 & 
0.000071 \\
  & $\text{MSE}_E$
    & {\bfseries 0.000435} & 
0.002253 & 
0.000599 \\
  & $\overline{F}$ 
    & {\bfseries 0.999116} & 
0.997990 & 
0.997795 \\
  & $\sigma_F$
    & {\bfseries 0.002301} & 
0.002837 & 
0.002492 \\
  & $\overline{\Delta \text{Tr}}$ 
    & 0.004945 & 
{\bfseries 0.002288} & 
0.004418 \\
  & $\sigma_{\Delta \text{Tr}}$
    & 0.003919 & 
0.002902 & 
{\bfseries 0.002813} \\
  & $\Delta t$(ns)
    & $-45.9\pm24.1$ & 
$-112.0\pm45.8$ & 
$\textbf{-4.9}\pm\textbf{26.8}$ \\
  & $\Delta \eta / \eta_{\text{true}}$(\%)
    & $\textbf{-3.83}\pm\textbf{7.32}$ & 
$71.24\pm59.76$ & 
$24.23\pm19.92$ \\
\bottomrule
\end{tabular}
\begin{tablenotes}
\item[a] See Table ~\ref{tab:evaluation_metrics} for metric definition.  
\end{tablenotes}
\end{threeparttable}




\end{appendices}


\bibliography{bibliography}


\begin{thebibliography}{39}
\ifx \bisbn   \undefined \def \bisbn  #1{ISBN #1}\fi
\ifx \binits  \undefined \def \binits#1{#1}\fi
\ifx \bauthor  \undefined \def \bauthor#1{#1}\fi
\ifx \batitle  \undefined \def \batitle#1{#1}\fi
\ifx \bjtitle  \undefined \def \bjtitle#1{#1}\fi
\ifx \bvolume  \undefined \def \bvolume#1{\textbf{#1}}\fi
\ifx \byear  \undefined \def \byear#1{#1}\fi
\ifx \bissue  \undefined \def \bissue#1{#1}\fi
\ifx \bfpage  \undefined \def \bfpage#1{#1}\fi
\ifx \blpage  \undefined \def \blpage #1{#1}\fi
\ifx \burl  \undefined \def \burl#1{\textsf{#1}}\fi
\ifx \doiurl  \undefined \def \doiurl#1{\url{https://doi.org/#1}}\fi
\ifx \betal  \undefined \def \betal{\textit{et al.}}\fi
\ifx \binstitute  \undefined \def \binstitute#1{#1}\fi
\ifx \binstitutionaled  \undefined \def \binstitutionaled#1{#1}\fi
\ifx \bctitle  \undefined \def \bctitle#1{#1}\fi
\ifx \beditor  \undefined \def \beditor#1{#1}\fi
\ifx \bpublisher  \undefined \def \bpublisher#1{#1}\fi
\ifx \bbtitle  \undefined \def \bbtitle#1{#1}\fi
\ifx \bedition  \undefined \def \bedition#1{#1}\fi
\ifx \bseriesno  \undefined \def \bseriesno#1{#1}\fi
\ifx \blocation  \undefined \def \blocation#1{#1}\fi
\ifx \bsertitle  \undefined \def \bsertitle#1{#1}\fi
\ifx \bsnm \undefined \def \bsnm#1{#1}\fi
\ifx \bsuffix \undefined \def \bsuffix#1{#1}\fi
\ifx \bparticle \undefined \def \bparticle#1{#1}\fi
\ifx \barticle \undefined \def \barticle#1{#1}\fi
\bibcommenthead
\ifx \bconfdate \undefined \def \bconfdate #1{#1}\fi
\ifx \botherref \undefined \def \botherref #1{#1}\fi
\ifx \url \undefined \def \url#1{\textsf{#1}}\fi
\ifx \bchapter \undefined \def \bchapter#1{#1}\fi
\ifx \bbook \undefined \def \bbook#1{#1}\fi
\ifx \bcomment \undefined \def \bcomment#1{#1}\fi
\ifx \oauthor \undefined \def \oauthor#1{#1}\fi
\ifx \citeauthoryear \undefined \def \citeauthoryear#1{#1}\fi
\ifx \endbibitem  \undefined \def \endbibitem {}\fi
\ifx \bconflocation  \undefined \def \bconflocation#1{#1}\fi
\ifx \arxivurl  \undefined \def \arxivurl#1{\textsf{#1}}\fi
\csname PreBibitemsHook\endcsname

\bibitem[\protect\citeauthoryear{Radnaev et~al.}{2010}]{radnaev2010quantum}
\begin{barticle}
\bauthor{\bsnm{Radnaev}, \binits{A.}},
\bauthor{\bsnm{Dudin}, \binits{Y.}},
\bauthor{\bsnm{Zhao}, \binits{R.}},
\bauthor{\bsnm{Jen}, \binits{H.}},
\bauthor{\bsnm{Jenkins}, \binits{S.}},
\bauthor{\bsnm{Kuzmich}, \binits{A.}},
\bauthor{\bsnm{Kennedy}, \binits{T.}}:
\batitle{A quantum memory with telecom-wavelength conversion}.
\bjtitle{Nature Physics}
\bvolume{6}(\bissue{11}),
\bfpage{894}--\blpage{899}
(\byear{2010})
\end{barticle}
\endbibitem

\bibitem[\protect\citeauthoryear{Wang et~al.}{2019}]{wang2019efficient}
\begin{barticle}
\bauthor{\bsnm{Wang}, \binits{Y.}},
\bauthor{\bsnm{Li}, \binits{J.}},
\bauthor{\bsnm{Zhang}, \binits{S.}},
\bauthor{\bsnm{Su}, \binits{K.}},
\bauthor{\bsnm{Zhou}, \binits{Y.}},
\bauthor{\bsnm{Liao}, \binits{K.}},
\bauthor{\bsnm{Du}, \binits{S.}},
\bauthor{\bsnm{Yan}, \binits{H.}},
\bauthor{\bsnm{Zhu}, \binits{S.-L.}}:
\batitle{Efficient quantum memory for single-photon polarization qubits}.
\bjtitle{Nature Photonics}
\bvolume{13}(\bissue{5}),
\bfpage{346}--\blpage{351}
(\byear{2019})
\end{barticle}
\endbibitem

\bibitem[\protect\citeauthoryear{Vernaz-Gris et~al.}{2018}]{vernaz2018highly}
\begin{barticle}
\bauthor{\bsnm{Vernaz-Gris}, \binits{P.}},
\bauthor{\bsnm{Huang}, \binits{K.}},
\bauthor{\bsnm{Cao}, \binits{M.}},
\bauthor{\bsnm{Sheremet}, \binits{A.S.}},
\bauthor{\bsnm{Laurat}, \binits{J.}}:
\batitle{Highly-efficient quantum memory for polarization qubits in a spatially-multiplexed cold atomic ensemble}.
\bjtitle{Nature communications}
\bvolume{9}(\bissue{1}),
\bfpage{363}
(\byear{2018})
\end{barticle}
\endbibitem

\bibitem[\protect\citeauthoryear{Chang et~al.}{2018}]{chang2018colloquium}
\begin{barticle}
\bauthor{\bsnm{Chang}, \binits{D.}},
\bauthor{\bsnm{Douglas}, \binits{J.}},
\bauthor{\bsnm{Gonz{\'a}lez-Tudela}, \binits{A.}},
\bauthor{\bsnm{Hung}, \binits{C.-L.}},
\bauthor{\bsnm{Kimble}, \binits{H.}}:
\batitle{Colloquium: Quantum matter built from nanoscopic lattices of atoms and photons}.
\bjtitle{Reviews of Modern Physics}
\bvolume{90}(\bissue{3}),
\bfpage{031002}
(\byear{2018})
\end{barticle}
\endbibitem

\bibitem[\protect\citeauthoryear{Reitz et~al.}{2022}]{reitz2022cooperative}
\begin{barticle}
\bauthor{\bsnm{Reitz}, \binits{M.}},
\bauthor{\bsnm{Sommer}, \binits{C.}},
\bauthor{\bsnm{Genes}, \binits{C.}}:
\batitle{Cooperative quantum phenomena in light-matter platforms}.
\bjtitle{Prx Quantum}
\bvolume{3}(\bissue{1}),
\bfpage{010201}
(\byear{2022})
\end{barticle}
\endbibitem

\bibitem[\protect\citeauthoryear{Masson and Asenjo-Garcia}{2022}]{masson2022universality}
\begin{barticle}
\bauthor{\bsnm{Masson}, \binits{S.J.}},
\bauthor{\bsnm{Asenjo-Garcia}, \binits{A.}}:
\batitle{Universality of dicke superradiance in arrays of quantum emitters}.
\bjtitle{Nature Communications}
\bvolume{13}(\bissue{1}),
\bfpage{2285}
(\byear{2022})
\end{barticle}
\endbibitem

\bibitem[\protect\citeauthoryear{Willis et~al.}{2009}]{willis2009four}
\begin{barticle}
\bauthor{\bsnm{Willis}, \binits{R.}},
\bauthor{\bsnm{Becerra}, \binits{F.}},
\bauthor{\bsnm{Orozco}, \binits{L.}},
\bauthor{\bsnm{Rolston}, \binits{S.}}:
\batitle{Four-wave mixing in the diamond configuration in an atomic vapor}.
\bjtitle{Physical Review A—Atomic, Molecular, and Optical Physics}
\bvolume{79}(\bissue{3}),
\bfpage{033814}
(\byear{2009})
\end{barticle}
\endbibitem

\bibitem[\protect\citeauthoryear{Ogden et~al.}{2019}]{ogden2019quasisimultons}
\begin{barticle}
\bauthor{\bsnm{Ogden}, \binits{T.P.}},
\bauthor{\bsnm{Whittaker}, \binits{K.}},
\bauthor{\bsnm{Keaveney}, \binits{J.}},
\bauthor{\bsnm{Wrathmall}, \binits{S.}},
\bauthor{\bsnm{Adams}, \binits{C.}},
\bauthor{\bsnm{Potvliege}, \binits{R.}}:
\batitle{Quasisimultons in thermal atomic vapors}.
\bjtitle{Physical Review Letters}
\bvolume{123}(\bissue{24}),
\bfpage{243604}
(\byear{2019})
\end{barticle}
\endbibitem

\bibitem[\protect\citeauthoryear{Corzo et~al.}{2019}]{corzo2019waveguide}
\begin{barticle}
\bauthor{\bsnm{Corzo}, \binits{N.V.}},
\bauthor{\bsnm{Raskop}, \binits{J.}},
\bauthor{\bsnm{Chandra}, \binits{A.}},
\bauthor{\bsnm{Sheremet}, \binits{A.S.}},
\bauthor{\bsnm{Gouraud}, \binits{B.}},
\bauthor{\bsnm{Laurat}, \binits{J.}}:
\batitle{Waveguide-coupled single collective excitation of atomic arrays}.
\bjtitle{Nature}
\bvolume{566}(\bissue{7744}),
\bfpage{359}--\blpage{362}
(\byear{2019})
\end{barticle}
\endbibitem

\bibitem[\protect\citeauthoryear{Kumar et~al.}{2023}]{kumar2023quantum}
\begin{barticle}
\bauthor{\bsnm{Kumar}, \binits{A.}},
\bauthor{\bsnm{Suleymanzade}, \binits{A.}},
\bauthor{\bsnm{Stone}, \binits{M.}},
\bauthor{\bsnm{Taneja}, \binits{L.}},
\bauthor{\bsnm{Anferov}, \binits{A.}},
\bauthor{\bsnm{Schuster}, \binits{D.I.}},
\bauthor{\bsnm{Simon}, \binits{J.}}:
\batitle{Quantum-enabled millimetre wave to optical transduction using neutral atoms}.
\bjtitle{Nature}
\bvolume{615}(\bissue{7953}),
\bfpage{614}--\blpage{619}
(\byear{2023})
\end{barticle}
\endbibitem

\bibitem[\protect\citeauthoryear{Fleischhauer and Lukin}{2000}]{fleischhauer2000dark}
\begin{barticle}
\bauthor{\bsnm{Fleischhauer}, \binits{M.}},
\bauthor{\bsnm{Lukin}, \binits{M.D.}}:
\batitle{Dark-state polaritons in electromagnetically induced transparency}.
\bjtitle{Physical review letters}
\bvolume{84}(\bissue{22}),
\bfpage{5094}
(\byear{2000})
\end{barticle}
\endbibitem

\bibitem[\protect\citeauthoryear{Fleischhauer and Lukin}{2002}]{fleischhauer2002quantum}
\begin{barticle}
\bauthor{\bsnm{Fleischhauer}, \binits{M.}},
\bauthor{\bsnm{Lukin}, \binits{M.D.}}:
\batitle{Quantum memory for photons: Dark-state polaritons}.
\bjtitle{Physical Review A}
\bvolume{65}(\bissue{2}),
\bfpage{022314}
(\byear{2002})
\end{barticle}
\endbibitem

\bibitem[\protect\citeauthoryear{Fleischhauer et~al.}{2005}]{fleischhauer2005electromagnetically}
\begin{barticle}
\bauthor{\bsnm{Fleischhauer}, \binits{M.}},
\bauthor{\bsnm{Imamoglu}, \binits{A.}},
\bauthor{\bsnm{Marangos}, \binits{J.P.}}:
\batitle{Electromagnetically induced transparency: Optics in coherent media}.
\bjtitle{Reviews of modern physics}
\bvolume{77}(\bissue{2}),
\bfpage{633}--\blpage{673}
(\byear{2005})
\end{barticle}
\endbibitem

\bibitem[\protect\citeauthoryear{Lvovsky et~al.}{2009}]{lvovsky2009optical}
\begin{barticle}
\bauthor{\bsnm{Lvovsky}, \binits{A.I.}},
\bauthor{\bsnm{Sanders}, \binits{B.C.}},
\bauthor{\bsnm{Tittel}, \binits{W.}}:
\batitle{Optical quantum memory}.
\bjtitle{Nature photonics}
\bvolume{3}(\bissue{12}),
\bfpage{706}--\blpage{714}
(\byear{2009})
\end{barticle}
\endbibitem

\bibitem[\protect\citeauthoryear{Namazi et~al.}{2017}]{namazi2017ultralow}
\begin{barticle}
\bauthor{\bsnm{Namazi}, \binits{M.}},
\bauthor{\bsnm{Kupchak}, \binits{C.}},
\bauthor{\bsnm{Jordaan}, \binits{B.}},
\bauthor{\bsnm{Shahrokhshahi}, \binits{R.}},
\bauthor{\bsnm{Figueroa}, \binits{E.}}:
\batitle{Ultralow-noise room-temperature quantum memory for polarization qubits}.
\bjtitle{Physical Review Applied}
\bvolume{8}(\bissue{3}),
\bfpage{034023}
(\byear{2017})
\end{barticle}
\endbibitem

\bibitem[\protect\citeauthoryear{Chaneli{\`e}re et~al.}{2005}]{chaneliere2005storage}
\begin{barticle}
\bauthor{\bsnm{Chaneli{\`e}re}, \binits{T.}},
\bauthor{\bsnm{Matsukevich}, \binits{D.}},
\bauthor{\bsnm{Jenkins}, \binits{S.}},
\bauthor{\bsnm{Lan}, \binits{S.-Y.}},
\bauthor{\bsnm{Kennedy}, \binits{T.}},
\bauthor{\bsnm{Kuzmich}, \binits{A.}}:
\batitle{Storage and retrieval of single photons transmitted between remote quantum memories}.
\bjtitle{Nature}
\bvolume{438}(\bissue{7069}),
\bfpage{833}--\blpage{836}
(\byear{2005})
\end{barticle}
\endbibitem

\bibitem[\protect\citeauthoryear{Gera et~al.}{2024}]{gera2024hong}
\begin{barticle}
\bauthor{\bsnm{Gera}, \binits{S.}},
\bauthor{\bsnm{Wallace}, \binits{C.}},
\bauthor{\bsnm{Flament}, \binits{M.}},
\bauthor{\bsnm{Scriminich}, \binits{A.}},
\bauthor{\bsnm{Namazi}, \binits{M.}},
\bauthor{\bsnm{Kim}, \binits{Y.}},
\bauthor{\bsnm{Sagona-Stophel}, \binits{S.}},
\bauthor{\bsnm{Vallone}, \binits{G.}},
\bauthor{\bsnm{Villoresi}, \binits{P.}},
\bauthor{\bsnm{Figueroa}, \binits{E.}}:
\batitle{Hong-ou-mandel interference of single-photon-level pulses stored in independent room-temperature quantum memories}.
\bjtitle{npj Quantum Information}
\bvolume{10}(\bissue{1}),
\bfpage{10}
(\byear{2024})
\end{barticle}
\endbibitem

\bibitem[\protect\citeauthoryear{Potvliege and Wrathmall}{2025}]{potvliege2025coombe}
\begin{barticle}
\bauthor{\bsnm{Potvliege}, \binits{R.}},
\bauthor{\bsnm{Wrathmall}, \binits{S.}}:
\batitle{Coombe: A suite of open-source programs for the integration of the optical bloch equations and maxwell-bloch equations}.
\bjtitle{Computer Physics Communications}
\bvolume{306},
\bfpage{109374}
(\byear{2025})
\end{barticle}
\endbibitem

\bibitem[\protect\citeauthoryear{Raissi et~al.}{2019}]{raissi2019physics}
\begin{barticle}
\bauthor{\bsnm{Raissi}, \binits{M.}},
\bauthor{\bsnm{Perdikaris}, \binits{P.}},
\bauthor{\bsnm{Karniadakis}, \binits{G.E.}}:
\batitle{Physics-informed neural networks: A deep learning framework for solving forward and inverse problems involving nonlinear partial differential equations}.
\bjtitle{Journal of Computational physics}
\bvolume{378},
\bfpage{686}--\blpage{707}
(\byear{2019})
\end{barticle}
\endbibitem

\bibitem[\protect\citeauthoryear{Bar-Sinai et~al.}{2019}]{bar2019learning}
\begin{barticle}
\bauthor{\bsnm{Bar-Sinai}, \binits{Y.}},
\bauthor{\bsnm{Hoyer}, \binits{S.}},
\bauthor{\bsnm{Hickey}, \binits{J.}},
\bauthor{\bsnm{Brenner}, \binits{M.P.}}:
\batitle{Learning data-driven discretizations for partial differential equations}.
\bjtitle{Proceedings of the National Academy of Sciences}
\bvolume{116}(\bissue{31}),
\bfpage{15344}--\blpage{15349}
(\byear{2019})
\end{barticle}
\endbibitem

\bibitem[\protect\citeauthoryear{Kovachki et~al.}{2023}]{kovachki2023neural}
\begin{barticle}
\bauthor{\bsnm{Kovachki}, \binits{N.}},
\bauthor{\bsnm{Li}, \binits{Z.}},
\bauthor{\bsnm{Liu}, \binits{B.}},
\bauthor{\bsnm{Azizzadenesheli}, \binits{K.}},
\bauthor{\bsnm{Bhattacharya}, \binits{K.}},
\bauthor{\bsnm{Stuart}, \binits{A.}},
\bauthor{\bsnm{Anandkumar}, \binits{A.}}:
\batitle{Neural operator: Learning maps between function spaces with applications to pdes}.
\bjtitle{Journal of Machine Learning Research}
\bvolume{24}(\bissue{89}),
\bfpage{1}--\blpage{97}
(\byear{2023})
\end{barticle}
\endbibitem

\bibitem[\protect\citeauthoryear{Lin et~al.}{2025}]{lin2025ai}
\begin{barticle}
\bauthor{\bsnm{Lin}, \binits{R.}},
\bauthor{\bsnm{Zhong}, \binits{H.-S.}},
\bauthor{\bsnm{Li}, \binits{Y.}},
\bauthor{\bsnm{Zhao}, \binits{Z.-R.}},
\bauthor{\bsnm{Zheng}, \binits{L.-T.}},
\bauthor{\bsnm{Hu}, \binits{T.-R.}},
\bauthor{\bsnm{Wu}, \binits{H.-M.}},
\bauthor{\bsnm{Wu}, \binits{Z.}},
\bauthor{\bsnm{Ma}, \binits{W.-J.}},
\bauthor{\bsnm{Gao}, \binits{Y.}}, \betal:
\batitle{Ai-enabled parallel assembly of thousands of defect-free neutral atom arrays}.
\bjtitle{Physical Review Letters}
\bvolume{135}(\bissue{6}),
\bfpage{060602}
(\byear{2025})
\end{barticle}
\endbibitem

\bibitem[\protect\citeauthoryear{von Glehn et~al.}{2022}]{von2022self}
\begin{botherref}
\oauthor{\bsnm{Glehn}, \binits{I.}},
\oauthor{\bsnm{Spencer}, \binits{J.S.}},
\oauthor{\bsnm{Pfau}, \binits{D.}}:
A self-attention ansatz for ab-initio quantum chemistry.
arXiv preprint arXiv:2211.13672
(2022)
\end{botherref}
\endbibitem

\bibitem[\protect\citeauthoryear{Choi et~al.}{2022}]{choi2022learning}
\begin{barticle}
\bauthor{\bsnm{Choi}, \binits{M.}},
\bauthor{\bsnm{Flam-Shepherd}, \binits{D.}},
\bauthor{\bsnm{Kyaw}, \binits{T.H.}},
\bauthor{\bsnm{Aspuru-Guzik}, \binits{A.}}:
\batitle{Learning quantum dynamics with latent neural ordinary differential equations}.
\bjtitle{Physical Review A}
\bvolume{105}(\bissue{4}),
\bfpage{042403}
(\byear{2022})
\end{barticle}
\endbibitem

\bibitem[\protect\citeauthoryear{Viteritti et~al.}{2025}]{viteritti2025transformer}
\begin{barticle}
\bauthor{\bsnm{Viteritti}, \binits{L.L.}},
\bauthor{\bsnm{Rende}, \binits{R.}},
\bauthor{\bsnm{Parola}, \binits{A.}},
\bauthor{\bsnm{Goldt}, \binits{S.}},
\bauthor{\bsnm{Becca}, \binits{F.}}:
\batitle{Transformer wave function for two dimensional frustrated magnets: Emergence of a spin-liquid phase in the shastry-sutherland model}.
\bjtitle{Physical Review B}
\bvolume{111}(\bissue{13}),
\bfpage{134411}
(\byear{2025})
\end{barticle}
\endbibitem

\bibitem[\protect\citeauthoryear{Zhang et~al.}{2025}]{zhang2025neural}
\begin{barticle}
\bauthor{\bsnm{Zhang}, \binits{J.}},
\bauthor{\bsnm{Benavides-Riveros}, \binits{C.L.}},
\bauthor{\bsnm{Chen}, \binits{L.}}:
\batitle{Neural quantum propagators for driven-dissipative quantum dynamics}.
\bjtitle{Physical Review Research}
\bvolume{7}(\bissue{1}),
\bfpage{012013}
(\byear{2025})
\end{barticle}
\endbibitem

\bibitem[\protect\citeauthoryear{Vaswani et~al.}{2017}]{vaswani2017attention}
\begin{botherref}
\oauthor{\bsnm{Vaswani}, \binits{A.}},
\oauthor{\bsnm{Shazeer}, \binits{N.}},
\oauthor{\bsnm{Parmar}, \binits{N.}},
\oauthor{\bsnm{Uszkoreit}, \binits{J.}},
\oauthor{\bsnm{Jones}, \binits{L.}},
\oauthor{\bsnm{Gomez}, \binits{A.N.}},
\oauthor{\bsnm{Kaiser}, \binits{{\L}.}},
\oauthor{\bsnm{Polosukhin}, \binits{I.}}:
Attention is all you need.
Advances in neural information processing systems
\textbf{30}
(2017)
\end{botherref}
\endbibitem

\bibitem[\protect\citeauthoryear{Dosovitskiy et~al.}{2020}]{dosovitskiy2020image}
\begin{botherref}
\oauthor{\bsnm{Dosovitskiy}, \binits{A.}},
\oauthor{\bsnm{Beyer}, \binits{L.}},
\oauthor{\bsnm{Kolesnikov}, \binits{A.}},
\oauthor{\bsnm{Weissenborn}, \binits{D.}},
\oauthor{\bsnm{Zhai}, \binits{X.}},
\oauthor{\bsnm{Unterthiner}, \binits{T.}},
\oauthor{\bsnm{Dehghani}, \binits{M.}},
\oauthor{\bsnm{Minderer}, \binits{M.}},
\oauthor{\bsnm{Heigold}, \binits{G.}},
\oauthor{\bsnm{Gelly}, \binits{S.}}, et al.:
An image is worth 16x16 words: Transformers for image recognition at scale.
arXiv preprint arXiv:2010.11929
(2020)
\end{botherref}
\endbibitem

\bibitem[\protect\citeauthoryear{Geneva and Zabaras}{2022}]{geneva2022transformers}
\begin{barticle}
\bauthor{\bsnm{Geneva}, \binits{N.}},
\bauthor{\bsnm{Zabaras}, \binits{N.}}:
\batitle{Transformers for modeling physical systems}.
\bjtitle{Neural Networks}
\bvolume{146},
\bfpage{272}--\blpage{289}
(\byear{2022})
\end{barticle}
\endbibitem

\bibitem[\protect\citeauthoryear{McCabe et~al.}{2024}]{mccabe2024multiple}
\begin{barticle}
\bauthor{\bsnm{McCabe}, \binits{M.}},
\bauthor{\bsnm{R{\'e}galdo-Saint~Blancard}, \binits{B.}},
\bauthor{\bsnm{Parker}, \binits{L.}},
\bauthor{\bsnm{Ohana}, \binits{R.}},
\bauthor{\bsnm{Cranmer}, \binits{M.}},
\bauthor{\bsnm{Bietti}, \binits{A.}},
\bauthor{\bsnm{Eickenberg}, \binits{M.}},
\bauthor{\bsnm{Golkar}, \binits{S.}},
\bauthor{\bsnm{Krawezik}, \binits{G.}},
\bauthor{\bsnm{Lanusse}, \binits{F.}}, \betal:
\batitle{Multiple physics pretraining for spatiotemporal surrogate models}.
\bjtitle{Advances in Neural Information Processing Systems}
\bvolume{37},
\bfpage{119301}--\blpage{119335}
(\byear{2024})
\end{barticle}
\endbibitem

\bibitem[\protect\citeauthoryear{Sprague and Czischek}{2024}]{sprague2024variational}
\begin{barticle}
\bauthor{\bsnm{Sprague}, \binits{K.}},
\bauthor{\bsnm{Czischek}, \binits{S.}}:
\batitle{Variational monte carlo with large patched transformers}.
\bjtitle{Communications Physics}
\bvolume{7}(\bissue{1}),
\bfpage{90}
(\byear{2024})
\end{barticle}
\endbibitem

\bibitem[\protect\citeauthoryear{Ho et~al.}{2019}]{ho2019axial}
\begin{botherref}
\oauthor{\bsnm{Ho}, \binits{J.}},
\oauthor{\bsnm{Kalchbrenner}, \binits{N.}},
\oauthor{\bsnm{Weissenborn}, \binits{D.}},
\oauthor{\bsnm{Salimans}, \binits{T.}}:
Axial attention in multidimensional transformers.
arXiv preprint arXiv:1912.12180
(2019)
\end{botherref}
\endbibitem

\bibitem[\protect\citeauthoryear{Liu et~al.}{2021}]{liu2021swin}
\begin{bchapter}
\bauthor{\bsnm{Liu}, \binits{Z.}},
\bauthor{\bsnm{Lin}, \binits{Y.}},
\bauthor{\bsnm{Cao}, \binits{Y.}},
\bauthor{\bsnm{Hu}, \binits{H.}},
\bauthor{\bsnm{Wei}, \binits{Y.}},
\bauthor{\bsnm{Zhang}, \binits{Z.}},
\bauthor{\bsnm{Lin}, \binits{S.}},
\bauthor{\bsnm{Guo}, \binits{B.}}:
\bctitle{Swin transformer: Hierarchical vision transformer using shifted windows}.
In: \bbtitle{Proceedings of the IEEE/CVF International Conference on Computer Vision},
pp. \bfpage{10012}--\blpage{10022}
(\byear{2021})
\end{bchapter}
\endbibitem

\bibitem[\protect\citeauthoryear{Liu et~al.}{2022}]{liu2022swin}
\begin{bchapter}
\bauthor{\bsnm{Liu}, \binits{Z.}},
\bauthor{\bsnm{Hu}, \binits{H.}},
\bauthor{\bsnm{Lin}, \binits{Y.}},
\bauthor{\bsnm{Yao}, \binits{Z.}},
\bauthor{\bsnm{Xie}, \binits{Z.}},
\bauthor{\bsnm{Wei}, \binits{Y.}},
\bauthor{\bsnm{Ning}, \binits{J.}},
\bauthor{\bsnm{Cao}, \binits{Y.}},
\bauthor{\bsnm{Zhang}, \binits{Z.}},
\bauthor{\bsnm{Dong}, \binits{L.}}, \betal:
\bctitle{Swin transformer v2: Scaling up capacity and resolution}.
In: \bbtitle{Proceedings of the IEEE/CVF Conference on Computer Vision and Pattern Recognition},
pp. \bfpage{12009}--\blpage{12019}
(\byear{2022})
\end{bchapter}
\endbibitem

\bibitem[\protect\citeauthoryear{Gao et~al.}{2022}]{gao2022earthformer}
\begin{barticle}
\bauthor{\bsnm{Gao}, \binits{Z.}},
\bauthor{\bsnm{Shi}, \binits{X.}},
\bauthor{\bsnm{Wang}, \binits{H.}},
\bauthor{\bsnm{Zhu}, \binits{Y.}},
\bauthor{\bsnm{Wang}, \binits{Y.B.}},
\bauthor{\bsnm{Li}, \binits{M.}},
\bauthor{\bsnm{Yeung}, \binits{D.-Y.}}:
\batitle{Earthformer: Exploring space-time transformers for earth system forecasting}.
\bjtitle{Advances in Neural Information Processing Systems}
\bvolume{35},
\bfpage{25390}--\blpage{25403}
(\byear{2022})
\end{barticle}
\endbibitem

\bibitem[\protect\citeauthoryear{Coopmans et~al.}{2021}]{coopmans2021netsquid}
\begin{barticle}
\bauthor{\bsnm{Coopmans}, \binits{T.}},
\bauthor{\bsnm{Knegjens}, \binits{R.}},
\bauthor{\bsnm{Dahlberg}, \binits{A.}},
\bauthor{\bsnm{Maier}, \binits{D.}},
\bauthor{\bsnm{Nijsten}, \binits{L.}},
\bauthor{\bsnm{Oliveira~Filho}, \binits{J.}},
\bauthor{\bsnm{Papendrecht}, \binits{M.}},
\bauthor{\bsnm{Rabbie}, \binits{J.}},
\bauthor{\bsnm{Rozp{\k{e}}dek}, \binits{F.}},
\bauthor{\bsnm{Skrzypczyk}, \binits{M.}}, \betal:
\batitle{Netsquid, a network simulator for quantum information using discrete events}.
\bjtitle{Communications Physics}
\bvolume{4}(\bissue{1}),
\bfpage{164}
(\byear{2021})
\end{barticle}
\endbibitem

\bibitem[\protect\citeauthoryear{Bartlett}{2018}]{bartlett2018distributed}
\begin{botherref}
\oauthor{\bsnm{Bartlett}, \binits{B.}}:
A distributed simulation framework for quantum networks and channels.
arXiv preprint arXiv:1808.07047
(2018)
\end{botherref}
\endbibitem

\bibitem[\protect\citeauthoryear{Wu et~al.}{2021}]{wu2021sequence}
\begin{barticle}
\bauthor{\bsnm{Wu}, \binits{X.}},
\bauthor{\bsnm{Kolar}, \binits{A.}},
\bauthor{\bsnm{Chung}, \binits{J.}},
\bauthor{\bsnm{Jin}, \binits{D.}},
\bauthor{\bsnm{Zhong}, \binits{T.}},
\bauthor{\bsnm{Kettimuthu}, \binits{R.}},
\bauthor{\bsnm{Suchara}, \binits{M.}}:
\batitle{Sequence: a customizable discrete-event simulator of quantum networks}.
\bjtitle{Quantum Science and Technology}
\bvolume{6}(\bissue{4}),
\bfpage{045027}
(\byear{2021})
\end{barticle}
\endbibitem

\bibitem[\protect\citeauthoryear{Lambert et~al.}{2024}]{lambert2024qutip5quantumtoolbox}
\begin{botherref}
\oauthor{\bsnm{Lambert}, \binits{N.}},
\oauthor{\bsnm{Giguère}, \binits{E.}},
\oauthor{\bsnm{Menczel}, \binits{P.}},
\oauthor{\bsnm{Li}, \binits{B.}},
\oauthor{\bsnm{Hopf}, \binits{P.}},
\oauthor{\bsnm{Suárez}, \binits{G.}},
\oauthor{\bsnm{Gali}, \binits{M.}},
\oauthor{\bsnm{Lishman}, \binits{J.}},
\oauthor{\bsnm{Gadhvi}, \binits{R.}},
\oauthor{\bsnm{Agarwal}, \binits{R.}},
\oauthor{\bsnm{Galicia}, \binits{A.}},
\oauthor{\bsnm{Shammah}, \binits{N.}},
\oauthor{\bsnm{Nation}, \binits{P.}},
\oauthor{\bsnm{Johansson}, \binits{J.R.}},
\oauthor{\bsnm{Ahmed}, \binits{S.}},
\oauthor{\bsnm{Cross}, \binits{S.}},
\oauthor{\bsnm{Pitchford}, \binits{A.}},
\oauthor{\bsnm{Nori}, \binits{F.}}:
Qutip 5: The quantum toolbox in python
(2024)
{\href{https://arxiv.org/abs/2412.04705}{{arXiv:2412.04705}}}
{[quant-ph]}
\end{botherref}
\endbibitem

\end{thebibliography}

\end{document}